\documentclass[smallextended]{svjour3}
\smartqed  
\usepackage{graphicx}
\usepackage{subfigure}
\usepackage[numbers,sort&compress]{natbib}
\usepackage{srcltx}
\usepackage{aas_macros}
\usepackage{authblk} 
\journalname{Experimental Astronomy}

\begin{document}
\title{The Large Observatory for X-ray Timing (LOFT)}


\author[,1,73]{M.~Feroci\footnote{marco.feroci@iasf-roma.inaf.it}}
\author[,2]{L.~Stella\footnote{luigi.stella@oa-roma.inaf.it}}
\author[3]{M.~van~der~Klis}
\author[4]{T.~J.-L.~Courvoisier}
\author[5]{M.~Hernanz}
\author[6]{R.~Hudec}
\author[7]{A.~Santangelo}
\author[8]{D.~Walton}
\author[9]{A.~Zdziarski}
\author[14]{D.~Barret}
% ...
\affil[1]{INAF/IASF, Via del Fosso del Cavaliere 100, I-00133 Roma, Italy} 
\affil[2]{INAF/OAR, Via di Frascati 33, I-00040 Monteporzio Catone, Italy} 
\affil[3]{Astronomical Institute Anton Pannekoek, University of Amsterdam, The Netherlands}
\affil[4]{ISDC, Geneva University, Switzerland}
\affil[5]{IEEC-CSIC, Spain} 
\affil[6]{Czech Technical University, Czech Republic} 
\affil[7]{Tuebingen University, Germany}
\affil[8]{MSSL-UCL, United Kingdom} 
\affil[9]{N. Copernicus Astronomical Center, Poland} 
% ...
\affil[14]{IRAP, France}
% ...
\affil[73]{INFN Roma Tor Vergata} 
% ...

\authorrunning{M. Feroci et al.} 

\institute{Marco Feroci for the LOFT team \at
              INAF/IASF, Via del Fosso del Cavaliere 100, I-00133 Roma, Italy\\
              Tel.: +39-06-4993 4099
              \email{marco.feroci@iasf-roma.inaf.it}
              }

\date{Received: date / Accepted: date}

\maketitle

\vspace{.5cm}

\begin{abstract}
High-time-resolution X-ray observations of compact objects provide
direct access to strong-field gravity, to the equation of state of
ultradense matter and to black hole masses and spins. A 10
m$^{2}$-class instrument in combination with good spectral
resolution is required to exploit the relevant diagnostics and
answer two of the fundamental questions of the European Space
Agency (ESA) Cosmic Vision Theme ``Matter under extreme
conditions'', namely: does matter orbiting close to the event
horizon follow the predictions of general relativity? What is the
equation of state of matter in neutron stars?
The Large Observatory For X-ray Timing (LOFT), selected by ESA as
one of the four Cosmic Vision M3 candidate missions to undergo an
assessment phase, will
  revolutionise the study of collapsed objects in our galaxy and of
  the brightest supermassive black holes in active galactic nuclei.
Thanks to an innovative design and the development of large-area
monolithic silicon drift detectors, the Large Area Detector (LAD)
on board LOFT will achieve an effective area of $\sim$12 m$^{2}$
(more than an order of magnitude larger than any spaceborne
predecessor) in the 2--30~keV range (up to 50~keV in expanded
mode), yet still fits a conventional platform and
small/medium-class launcher. With this large area and a spectral
resolution of $<$260 eV, LOFT will yield unprecedented information
on strongly curved spacetimes and matter under extreme conditions
of pressure and magnetic field strength. 


\PACS{TBI}

\end{abstract}

\section{Introduction}\label{s:introduction}
Our knowledge of compact objects derives from the discoveries made
with a steady progression of ever larger-area, hence more
sensitive, X-ray timing instrumentation, from Uhuru (US, 0.08
m$^2$, \cite{Giacconi1971}), via satellites such as EXOSAT (ESA,
\cite{Peacock1981}) and Ginga (Japan, \cite{Turner1989}) to, most
recently, RXTE (US, 0.6 m$^2$, \cite{Swank2006}). The latter
instrument, thanks to it having the largest area so far, finally
made good on the promise (e.g. \cite{Shvartsman1971}) that the
process of accretion onto stellar mass collapsed objects should
exhibit variability at the millisecond dynamical time scale of the
small (few Schwarzschild radii, i.e., kilometres) regions where
most of the gravitational potential energy is released.
Quasi-periodic oscillations (QPOs) at this time scale were found
in the X-ray flux of both neutron stars (kHz QPOs, up to 1250~Hz)
and black holes (high frequency QPOs, up to 450~Hz) indicating the
action of mechanisms in the strong field region picking out
specific frequencies, e.g., orbital and epicyclic motion at
specific radii near the marginally stable orbit ($r_\mathrm{isco}=
6\,GM/c^2$ for a Schwarzschild black hole) as predicted in General
Relativity. The long-sought millisecond spins in accreting neutron
stars were also finally found, both during thermonuclear bursts
(burst oscillations) and in the persistent flux (coherent periodic
signals of millisecond pulsars), as were the vibrational responses
to cataclysmic events of some neutron stars themselves (magnetar
crust/core oscillations). Relativistically broadened Fe K-lines
observed from the accretion disks of both X-ray binaries and
active galactic nuclei (AGNs) are thought to arise in the same
inner regions of the accretion flow. All these phenomena are
direct diagnostics of the motion of matter and the propagation of
radiation in very strong gravitational fields: they encode
information about the mass, radius and spin of collapsed objects
and can probe the properties of ultradense matter and superstrong
magnetic fields in neutron stars. Yet tantalizingly, while the
relevant signals were identified, due to its still limited
sensitivity, RXTE was only able to scratch the surface when it
came to actually using these diagnostics to probe strong field
gravity and ultradense matter.

LOFT is specifically designed to exploit the diagnostics of very
rapid X-ray flux and spectral variability that directly probe the
motion of matter down to distances very close to black holes and
neutron stars. Its $\sim$20 times larger effective area than
RXTE's PCA  is crucial in this respect. The key to this
breakthrough in effective area resides in the synergy between
technologies imported from other fields of scientific research.
The crucial characteristic of the LOFT Large Area Detector (LAD)
is a mass per unit surface in the range of $\sim$10 kg~m$^{-2}$,
enabling a payload with $\sim$15~m$^{2}$ geometric area at
reasonable weight. The ingredients for a sensitive but lightweight
experiment are the large-area Silicon Drift Detectors (SDDs)
designed on the heritage of the ALICE experiment at CERN/LHC
\cite{Vacchi1991}, and a collimator based on lead-glass capillary
plates. The drift concept makes the spectroscopic performance of
the SDDs weakly dependent on the extent of the collecting surface:
large-area ($\sim$70~cm$^{2}$) monolithic detectors can be
designed, with only 256 read-out anodes (thus low power
requirements), but very good spectral performance (FWHM
$\sim$260~eV) \cite{Zampa2011}.

An unprecedentedly large throughput ($\sim3 \times 10^{5}$
cts~s$^{-1}$ from the Crab) is achieved with a segmented detector,
making pile-up and dead-time, often worrying or limiting focused
experiments, secondary issues. The large detector on LOFT is
deployed in space through a mechanism relying on the experience in
Synthetic Aperture Radar missions (e.g., the SMOS ESA mission, in
orbit since 2009, \cite{Bueno2005}), where very large panels are
deployed in space with high accuracy. This mechanism, combined
with the low-weight detector technology, allows the stowing of the
LOFT satellite inside the fairing of a Vega launcher to design a
small mission in a low-background equatorial LEO ($\sim$600~km).
The LOFT scientific payload is completed by a coded-mask Wide
Field Monitor (WFM), that uses the same type of detectors, and is
in charge of monitoring a large fraction of the sky potentially
accessible to LAD, to provide the history and context for the
sources observed by LAD and to trigger its pointed observations on
their most interesting and extreme states. Its sensitivity and
large sky coverage make the WFM also an important resource on its
own right.

In this paper we describe the LOFT scientific objectives and
expected scientific performance
(Sect.~\ref{s:scientific_objectives}), mission profile
(Sect.~\ref{s:profile}) and details of the instrumentation and
model payload designed to achieve the science objectives
(Sect.\ref{s:model_payload}).

\section{Scientific objectives}\label{s:scientific_objectives}

\subsection{The properties of ultradense matter}\label{s:ultradense_matter}
One of the key goals of high-energy astrophysics is to determine
the equation of state (EOS) of ultradense matter, i.e. the
relation between density and pressure at the highest possible
densities, where the physics of quarks and quantum chromodynamics
(QCD) come into play. At densities exceeding a few times the
density of atomic nuclei, exotic states of matter such as Bose
condensates or hyperons may appear. At even higher densities a
phase transition to strange quark matter may also occur. This
high-density/low-temperature region of the QCD phase diagram is
inaccessible to terrestrial laboratories and can only be probed
astrophysically, where the ultradense matter EOS manifests as a
mass-radius (M-R) relation for neutron stars (NS), whose cores
contain the densest matter known. Many theoretical EOSs have been
proposed through the years, each predicting a different M-R
relation for neutron stars. Model EOSs may be classified by the
maximum neutron star mass that the EOS can sustain:  ``soft'' EOSs
(which produce neutron stars with low central density) have a
maximum mass in the 1.5--1.7 solar masses (M$_{\odot}$) range,
whereas ``stiff'' EOSs can reach up to 2.4--2.5~M$_{\odot}$
\cite{Lattimer2001}. Recently, a heavy neutron star mass of
$1.97\pm0.04$~M$_{\odot}$ has been measured through accurate
timing of a millisecond radio pulsar in a relativistic binary
\cite{Demorest2010}, ruling out most soft EOSs and limiting the
presence of quark matter in the neutron star core. This is an
important step, but the quest for the neutron star EOS is far from
settled. We still do not know the radius or maximum mass that a
star can support before collapsing into a black hole.

\subsubsection{Neutron star mass and radius measurements}\label{s:ns_eos}
Neutron star masses are measured with great precision from binary
radio pulsar timing. However, neutron star radius measurements are
particularly challenging, and no reliably precise measurements are
available to date. Even more challenging is the contemporaneous
measurement of $M$ and $R$, which would allow clear discrimination
between EOS models.

Several techniques to estimate $M$ and/or $R$ have been proposed,
discussed and used in the past \cite{Lattimer2007}. The best tool
to constrain the neutron star EOS is the modelling of coherent
X-ray pulsations, that are detected in accretion-powered
millisecond X-ray pulsars and during thermonuclear explosions on
the surface of many accreting neutron stars (``type I'' bursts).
The observed pulsed signal is shaped by the brightness pattern on
the stellar surface and by gravitational self-lensing by the
neutron star's strong gravity, allowing to view more than just the
facing hemisphere. This effect alters the amplitude of pulsations
generated by rotating ``hot spots'' on the surface, and it
strongly depends on the neutron star $M/R$ ratio. Similarly, the
shape of the pulse is affected by the stellar rotational velocity
(which depends on the star's radius and hot spot latitude) and by
the binary inclination. Modelling the variations of the pulse
shape in different energy bands and at different luminosity levels
has been demonstrated effective in constraining mass and radius
\cite{Leahy2009}. Simulations based on RXTE results show that LOFT
can achieve $\sim$5\% precision in radius measurement with such
observations. Effective discrimination between different families
of EOSs will be possible only with such a relative precision.

In $\sim$25 neutron stars, coherent oscillations at frequencies of
up to 620~Hz are also observed during type I X-ray bursts, which
are mostly the result of thermally unstable helium ignition in the
accreted envelope of a neutron star. This generates a
thermonuclear explosion that is observed as an X-ray burst with a
rapid rise ($\sim$1~s) followed by a slower decay
($\sim$10--100~s). X-ray pulsations can be measured with
sufficient precision during the burst rise (i.e. when the X-ray
emitting area grows) by a detector with very large effective area.
Simulations show that with LOFT we can achieve an error of less
than 5\% (90\% c.l.) on both $M$ and $R$, firmly constraining the
EOS.

In the X-ray spectra of two photospheric radius expansion bursts
observed with the RXTE satellite, despite the low spectral
resolution, strong absorption features were detected during the
early expansion phase with a timescale of $\sim$1~s
\cite{intZand2010}. These features are predicted due to the
presence of ashes of nuclear burning in the photosphere. The
changing edge energy during the burst precluded a measurement of
the gravitational redshift and the low resolution an unambiguous
identification of nickel as the element. Future measurements with
larger effective areas and moderately higher spectral resolution,
like provided with LOFT/LAD, will determine the redshift at the
neutron star surface.

Fast rotation of compact stars is sensitive to the stellar mass
and to the EOS, constraining the star spin against centrifugal
break-up. The fastest known neutron star spin (716 Hz, and, if
confirmed, 1122 Hz) already rules out an entire class of models
predicting large, low-mass neutron stars. This search in X-rays
has no known biases against the detection of very high spin rates
(unlike radio pulsation searches) and if carried out with a very
large area as LOFT's will achieve very high sensitivities: a
pulsed fraction of 0.8\% and 0.07\% will be detected at
5\,$\sigma$ with a 100~s observation of a 100~mCrab source and
Sco~X-1, respectively. The observed lack of extremely high spin
frequencies ($>$800~Hz) in the most rapidly accreting neutron
stars also suggests that their accretion torques might be balanced
by gravitational wave emission torques. LOFT will provide a
complete census of periods in X-ray binaries. Moreover a precise
knowledge of the spin periods of fast spinning neutron stars can
greatly increase the sensitivity of the searches for their
gravitational wave signals with interferometers such as Advanced
Virgo/LIGO.

\subsubsection{Neutron star crust properties}\label{s:ns_crust}
In 2004, the magnetar SGR~1806-20 emitted the most powerful flare
ever recorded, with a fluence of several 10$^{47}$ ergs (a.k.a.
giant flare), affecting even the Earth's ionosphere. QPOs
($\sim$10\% rms amplitude) in the 18--1800~Hz range were detected
in the X-ray band during the decaying tail of the flare
\cite{Israel2005}. Similar results were obtained for SGR~1900+14
by re-analyzing the X-ray archival data of its 1998 giant flare
\cite{Strohmayer2005}. Detection of signals at similar frequencies
in the giant flares from two different magnetars implies that the
same process is operating in both objects. Powering the rare giant
flares requires a catastrophic reconfiguration of the magnetic
field and likely implies large-scale crust fracturing (i.e.
starquakes). This in turn should excite coupled crust-core
(``seismic'') oscillations observed in the X-rays as QPOs.
Different types of these global seismic oscillation (GSOs) are
possible, torsional magneto-elastic modes in particular are
expected to give rise to the observed oscillations. The harmonics
which are excited depend on the size, shape and speed of the
fracture. The observed mode frequencies depend on the neutron star
mass and radius, crustal composition, core superconductor state
and magnetic field strength and configuration. The identification
of the GSOs modes discovered in the two giant flares remains a
subject of active theoretical debate. At present it seems possible
to explain them in terms of magnetic-elastic modes dominated
either by the magnetic field or by the crust, depending on the
state of the core superconductor.

\begin{figure}[t]
\centering
    \includegraphics[width=0.6\textwidth]{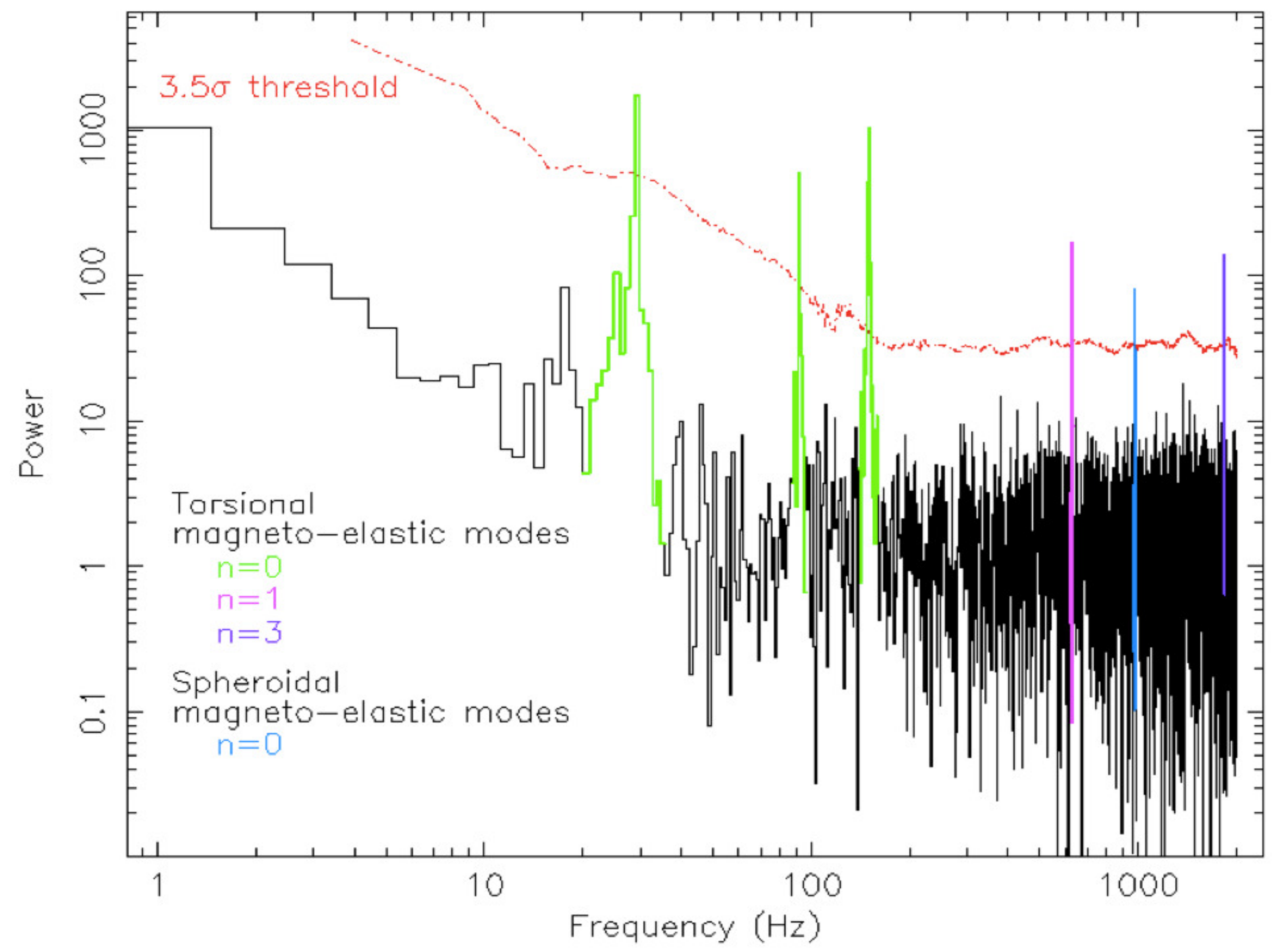}
    \caption{Frequencies expected in the power spectrum of a magnetar intermediate flare as seen by LOFT.}
    \label{f:intermediate_flare_PSD}
\end{figure}

Giant flares are rare and the chances that one of them occurs
during the operational lifetime of LOFT are fairly low (30\%). On
the other hand, theory tells us that GSOs should be excited also
during the much more frequent (tens during active states) but less
energetic events called intermediate flares (duration 1--60~s,
fluence up to 10$^{43}$ ergs). LOFT, thanks to its large area and
monitoring capabilities, will be able to detect and study in depth
magnetar QPOs during intermediate flares. This will be done on
bright and hard serendipitous events shining in the experiment
from directions outside the field of view (in this case the LAD
response will be mostly above $\sim$20--30~keV, by using the
extended energy range, where the collimator opacity decreases) or
by direct pointing of magnetars in their burst active state,
following a trigger by the WFM (in which case the events will be
covered with the whole effective area of the LAD down to energies
of 2~keV). Studying these events with LOFT's exceptionally high
throughput will detect much weaker magnetar QPOs than observed so
far, thus probing the interiors of neutron stars in a similar
manner to helioseismology. A 500 Crab intermediate flare shining
at 30$^{\circ}$ offset angle in the LAD collimators would allow
the detection of magnetar QPOs with amplitudes as low as 5\% and
0.7\%, respectively, for a flare duration of 1~s and 60~s (Fig.
\ref{f:intermediate_flare_PSD}).

Searches for seismic oscillations can be also carried out in
glitches observed in magnetars as well as in rotation-powered
radio pulsars. Glitches are thought to originate from a sudden
unpinning of angular momentum vortices in the neutron star crustal
superfluid. The recovery from a glitch contains information on the
neutron star interior and in particular on the coupling between
crust and core. Differences among the behaviour of radio pulsars
and magnetars can lead to the discovery of systematic structural
differences between the two types of neutron stars.
\subsection{Strong gravity and the mass and spin of black holes}\label{s:strong_gravity}
Most of the power generated by accretion onto neutron stars and
black holes is released deep in the gravitational potential wells
of these objects, where the matter flowing inwards moves at a
sizable fraction of the speed of light. Observation of the motion
of matter in the innermost regions of accretion disks contains
unique information on the properties of the strongly curved
spacetime near neutron stars and black holes, and on the compact
object's mass, spin and size. Over the last two decades powerful
diagnostics of the innermost disk regions have been discovered:
among those based on timing measurements are the QPOs in the flux
of a number of X-ray binary systems (and until now one active
galactic nucleus) which occur at the dynamical time scale of the
innermost disk regions and hence are associated to the fundamental
frequencies of matter motion in strong field gravity. The Fe
K$_{\alpha}$ emission lines which are observed in many accreting
collapsed objects, from neutron stars to supermassive black holes
in AGNs, are powerful spectral diagnostics that form a second,
independent probe of the same strong field gravity regions.

The exploitation of these diagnostics has so far been hampered by
insufficient throughput and/or energy range and resolution of the
instrumentation. For instance, in about 20 neutron star binary
systems, a diverse QPO phenomenology was observed with RXTE. At
high frequencies (200--1200~Hz), pairs of signals are observed,
whose frequencies vary on timescales as short as hundreds of
seconds. These QPOs can be strong, but their intensity is highly
variable. Averaging these signals over thousands of seconds,
sometimes days in order to detect them, has until now prevented
the exploitation of the full potential of this diagnostic. A
somewhat similar phenomenology is observed in several black holes,
but their QPOs are weaker and even more difficult to detect,
especially at high frequencies. The current indication is that the
frequencies of these high frequency QPOs are rather stable for
each system and anti-correlate with the black-hole mass. In a few
cases where two frequencies can be identified, they appear to be
at special ratios of 2:3 and 3:5. These features have been
detected rarely and only in a specific state when sources are
bright, nearly certainly because only then they peak up to levels
detectable with present instruments \cite{Vanderklis2000}.

\subsubsection{Neutron star and stellar mass black hole}\label{s:ns_and_stellar_bh}
The origin of QPOs in neutrons stars and black holes is still open
to different interpretations, but all viable explanations include
a hydrodynamical flow in which the fundamental frequencies of
motion (azimuthal, epicyclic) in the strong gravitational field
near these objects dominate the motion. In this region, GR effects
are important: for instance there is no post-Newtonian treatment
that could approximate the behaviour of the radial epicyclic
frequency in this regime. The magnitude of general relativistic
effects is thus not an issue. It is crucial instead to make sure
that the QPO diagnostics are interpreted correctly. The major leap
in the effective area that LOFT provides will permit for the first
time to exploit new, qualitatively different techniques for
extracting physical information from the QPOs. In the following we
describe a few examples of this.

The relative ``simplicity'' of the space-time around black holes
(cleared of the complications of solid surface, magnetosphere and
internal mass distribution effects on the external spacetime that
affect neutron stars) makes the theoretical modeling relatively
straightforward, thus the observational tests particularly
relevant and conclusive. Methods have been devised to test the
models through detailed high signal to noise observations. For
instance, in the parametric epicyclic resonance model two of the
oscillation modes are associated with a resonance of the
relativistic radial and vertical epicyclic frequencies
\cite{Abramowicz2001}; in the relativistic precession model
instead, the highest frequency signal is identified with the
azimuthal frequency of motion, while two of the modes arise from
relativistic nodal and periastron precession (whose frequency is
related also to the epicyclic frequencies \cite{Stella1999}). By
studying the QPO behaviour over a range of source intensities LOFT
will afford to remove the current degeneracy in the interpretation
of the oscillation.
Once the ambiguity in the
interpretation of the QPO phenomena is resolved, the frequencies
of the QPO themselves, which LOFT measures with exquisite
precision, will immediately provide access to yet unobserved
general relativistic effects such as Lense-Thirring and strong
field periastron precession, or the unique behaviour of the
epicyclic frequencies near a black hole. As these frequencies
embody also information on the black hole mass and spin, these can
thus be derived or, at least, tightly constrained.

LOFT will enable the study of the QPO signals in time (as opposed
to Fourier) domain; in particular, low frequency QPOs will be
studied in individual pulses and kHz QPO within their coherence
timescale. (The latter requires a minimum effective area of
$\sim$10~m$^2$). Time-domain studies will open a new dimension in
the exploitation of QPOs: it will permit to directly observe the
decoherence process and recurrence of these signals, measure
waveforms and their energy dependence, perform phase-resolved
spectroscopy, and compare these direct diagnostics of what is
going on at the relevant (for the high frequency QPOs: dynamical)
timescales with numerical work predicting these behaviours of the
accretion flow. This will be crucial to decide whether QPOs
originate from, e.g., orbiting hot spots and disk vibration or
warping modes.

In nearly all cases discussed above, additional information will
be derived from the time-averaged Fe K$_{\alpha}$ fluorescence
line profile. As is well known, relativistic effects such as
gravitational redshift, light-bending, framedragging, and Doppler
shifts distort the iron line shape into a characteristic broad
shape. Measurements of the line profile then translate into
measurements of the innermost radius of the disk (in natural units
of $r_g=GM/c^2$), the inclination and emissivity law of the
disk. Through the effects of frame dragging on the innermost
stable orbit, this leads to the determination of the spin of the
black hole. This spectroscopic measurement of the matter motions
near the central object can be combined with timing measurements
to derive compact object mass and orbital radii. The energy
resolution and large area of LOFT will allow measurements of
unprecedented quality for this relativistically broadened iron
line and its variations.

\begin{figure}[!t]
 \centering
  \includegraphics[width=0.6\textwidth]{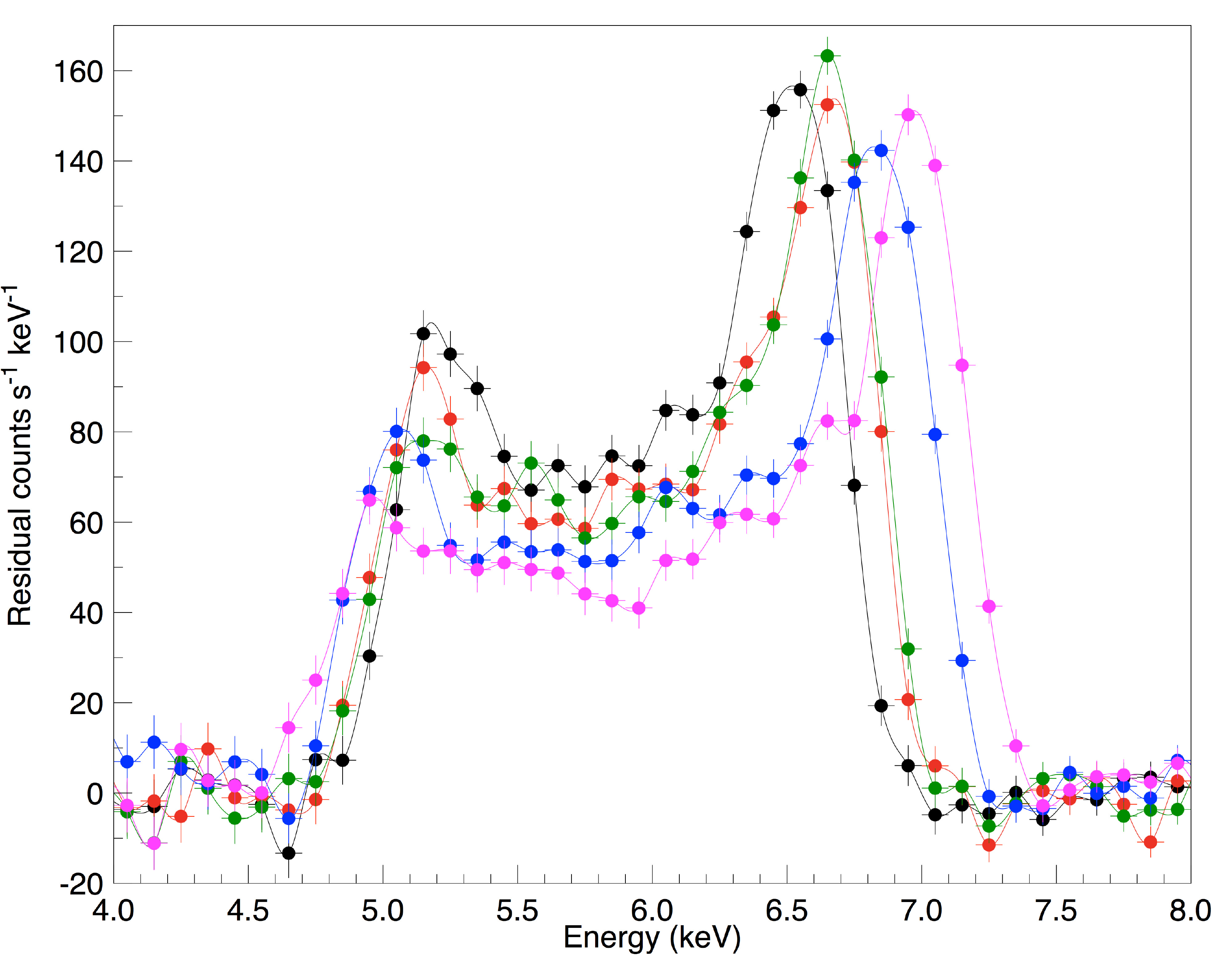}
   \caption{Contribution to the Fe line emission of a small ring (9--10 $r_{g}$)
   of matter undergoing nodal (Lense-Thirring) precession around a neutron star.
   The different line profiles are computed for different phase of the
   precessional motion; the tilt angle relative to the disk is 5$^{\circ}$
   around the system inclination (26$^{\circ}$). The simulation here is
   representative of  the case of the ~300 mCrab neutron star binary GX 349+2.
   The figure shows the contribution from the ring emission, after removal of
   the continuum and steady Fe-line components.}
\label{f:Fe_line_precession}
\end{figure}

We illustrate here an example of the potential of the combined
exploitation of flux and spectral variability with LOFT, under the
hypothesis that the low frequency QPO originate from the nodal
precession of the innermost disk region as envisaged in some QPO
models. The varying inclination angle caused by precession will
induce a characteristic variation in the width and profile of the
Fe K-line produced by the innermost disk region. Fig.
\ref{f:Fe_line_precession} shows the result of a simulation in
which the inner disk ring of the accretion disk undergoes a 30 Hz
nodal precession with a 5 degrees tilt angle relative to the rest
of the disk; the LAD will reveal the precession of the inner disk
with very high significance.

\subsubsection{Supermassive black holes}\label{s:smbh}
Bright AGNs provide an additional and somewhat complementary
opportunity to study strong field gravity. Though their flux at
the Earth is typically 100--1000 times lower than that of
accreting black holes in galactic binaries, their dynamical
timescales (which scale with the mass) are $\sim$10$^{6}$ times
longer. The result is that some 10$^{3}$ times more photons are
received at the Earth per dynamical timescale from bright AGNs
than from X-ray binaries. Therefore bright AGN provide a parallel
channel for investigating individual realizations of very short
term phenomena, such as the motion of a single spot in the disk or
the response of the Fe-line disk emissivity to flares from the
illuminating source.

As in X-ray binaries, the relativistic broad Fe-line profile that
is seen in the X-ray spectrum of a number of AGNs (about 70\% of
bright Seyfert~1), provides a powerful tool to probe the accretion
flow in a region where the motion is determined by GR
\cite{Fabian1989, Laor1991}. Evidence for spinning supermassive
black holes has been found in several cases. LOFT will determine
with very high signal to noise and accurate continuum subtraction
the profile of AGN Fe K-lines, thanks to its sensitivity and broad
energy range. The instrumental and cosmic background in the LAD
will permit to study AGNs in this context down to flux level of
$\sim$1~mCrab, with signal to noise ratios of about 500 in
10$^{4}$ s exposure. With a 10$^3$~s exposure, LOFT will collect
more than 5$\times$10$^5$~counts in 2--30 keV for a 3~mCrab AGN:
that will provide a high enough S/N to determine accurately the Fe
line profile and measure the inner radius of the disk down to the
marginally stable orbit and from this derive the spin of the BH.

\begin{figure}[!t]
 \centering
  \includegraphics[width=0.7\textwidth]{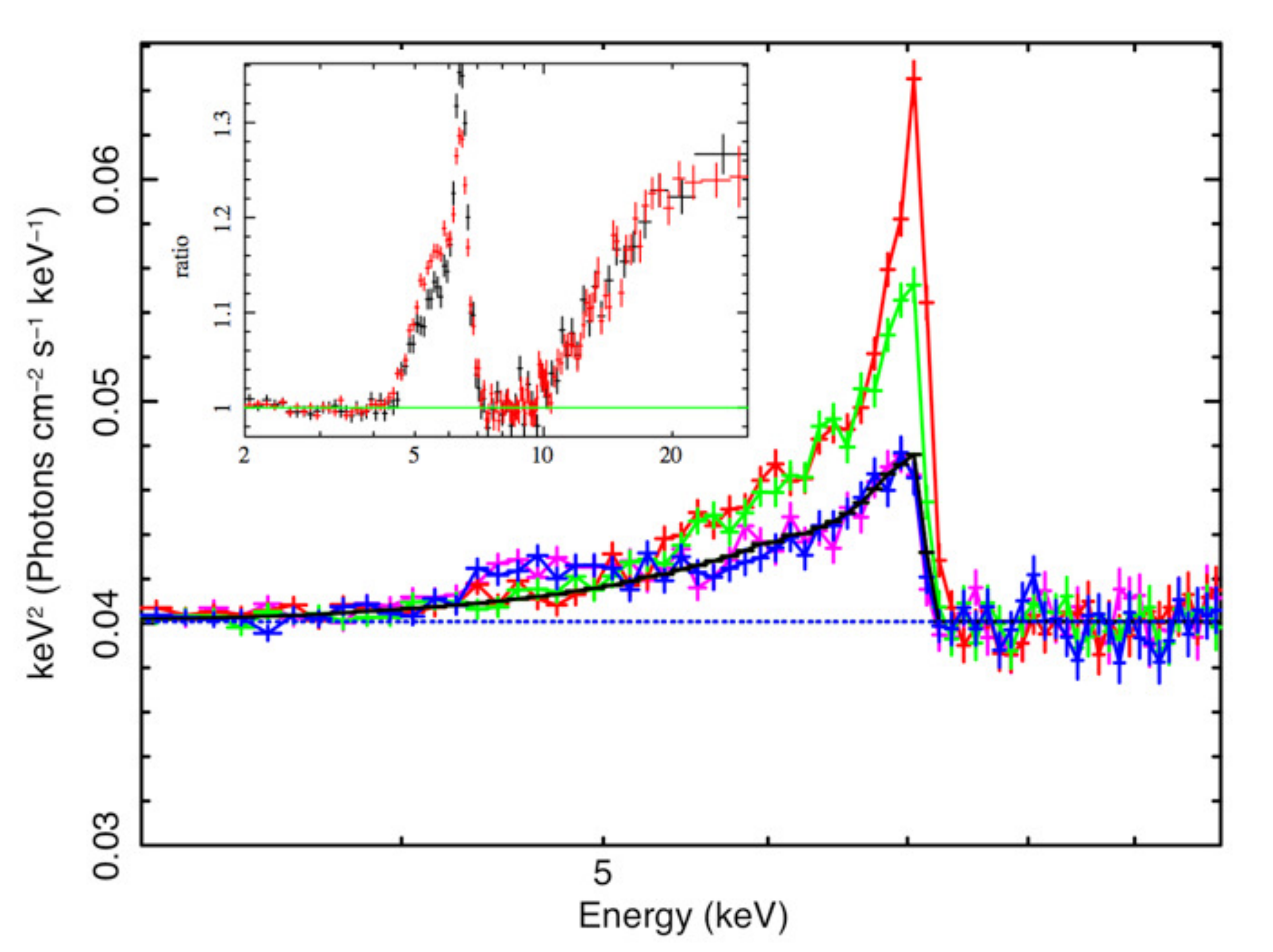}
   \caption{Simulated LOFT/LAD spectra of  a 3~mCrab (2--10 keV) AGN.
   The lower plot shows the broad relativistic
   Fe line produced in the innermost region of the accretion disc extending
   down to the marginally stable for an almost-maximally rotating black hole
   ($a=0.99$) with mass of 3.6 $\times$ 10$^{6}$ M$_\odot$, viewed at an inclination
   of 45 deg. The plot shows the variable Fe K feature produced by an orbiting
   spot at $r=10\,GM/c^{2}$ on the disk's surface that is illumination by a flare.
   The orbital period is 4 ks and the total exposure is 16 ks.
   LOFT/LAD will track the line variation on 1 ks time scale, thus allowing a
   determination of the orbital period through measurements at four different
   orbital phases over four cycles. The inset shows two 10 ks spectra from
   simulated high (6 mCrab in 2--10 keV) and low (2 mCrab) flux states LOFT/LAD
   observations of MCG-6-30-15. Thanks to the wide energy band and the large
   effective area above 10 keV, the variation of the reflection hump can be
   measured with 2\% accuracy. }
\label{f:Fe_line_AGN}
\end{figure}

Moreover, LOFT's very high throughput will permit to investigate
the line response to flares and reveal the orbital motion of
individual blobs illuminated by the central source through the
moving features that they induce in the line profile
\cite{Dovciak2004}. These ``reverberations'' will provide the
physical unit for the emitting radius that is required to measure
also the black hole mass in addition to the spin. An example of
the expected performance of LOFT on this subject is shown in Fig.
\ref{f:Fe_line_AGN}.

\subsection{Additional Science Objectives}\label{s:additional_science}
In addition to the main science goals outlined above, LOFT will
very successfully exploit the physical information contained in
the flux and spectral variability of many different classes of
sources to address a number of open astrophysical problems. The
longer timescale variability will be explored by the WFM, singling
out specific source activities and/or spectral states (e.g. those
of accreting millisecond transient pulsars) amongst many hundreds
of monitored sources. Clearly the WFM will be essential to trigger
and carry the LAD observations required to achieve LOFT's main
science objectives. In the following we briefly outline some of
additional science themes to which LOFT will give great
contributions.

\paragraph{X-ray binaries} are the brightest sources of the X-ray sky.
 LOFT will be able to detect periodicities of unprecedented small
amplitude and measure their phase and period evolution to high
accuracy.  By means of the study of periodic signals, several
issues could be investigated:
\begin{itemize}
    \item the physics and geometry of accretion onto magnetic white dwarfs and neutron stars;
     \item the torques responsible for their secular spin up and down and the instabilities giving rise to glitches;
    \item evidence for the long-sought photon bubble instability in the accretion column above magnetic neutron stars;
    \item intermittent millisecond pulsations in accreting X-ray binaries.
\end{itemize}

\paragraph{Millisecond radio pulsars} also emit in the X-ray band. A large number of $\gamma$-ray sources have been discovered with
Fermi, that are suspected to be fast rotation powered pulsars.
However, many of these are in regions of the Galaxy behind large
foreground columns of ionized gas and dust, implying that neither
radio, nor soft X-ray followup will be able to confirm their
nature; however, a large area hard X-ray timing mission like LOFT
should be able to detect the pulsations from many such objects.

\paragraph{The mechanism for the launching of jets} is still poorly understood. The WFM's excellent sensitivity  will allow us to
trigger early on new X-ray transients, and hence to use the LAD
for multi-wavelength campaigns in conjunction with optical and IR
instruments with sub-second time resolution (e.g. ISAAC on the
VLT, and SALTICAM on SALT). Cross-correlations between X-rays and
optical/IR  can be used to estimate jet speeds even in cases where
the spatial scale of the jets are too small to be resolved
spatially. By being able to make these measurements over a wide
range of luminosities for the same sources, it will be possible to
determine relations between jet speeds and luminosities and
accretion states.

\paragraph{WFM's large instantaneous field of view} is another
important advantage of LOFT over past soft X-ray all-sky
instruments. This gives LOFT an unprecedented capability for
detecting rare, short-lived, bright sources, such as Gamma Ray
Bursts, X-ray flashes, X-ray transients of all classes , and for
obtaining high cadence monitoring of moderately bright sources.
One of the RXTE highlights was the monitoring of many active
galactic nuclei using pointed observations that allowed to
demonstrate that many of these systems had power spectral breaks
similar to those seen in X-ray binaries. Similar monitoring, with
much better sampling, can be done on all similarly bright AGNs
with LOFT/WFM. Pointed observations with the LOFT/LAD will be able
to detect short term variations of $\sim$1--3 mCrab AGNs with
unprecedented sensitivity. Hard X-ray monitoring of flaring
blazars is especially important, since there is a strong
correlation between hard X-ray activity and high-energy gamma-ray
activity. LOFT's WFM can be used to trigger ground-based Cherenkov
telescopes, and comparisons of the synchrotron peak in the X-rays
with the Compton peak in the gamma-rays can be used to provide
constraints on the cosmic infrared background.

\section{Mission profile} \label{s:profile}
\begin{table}
\begin{center}
\begin{tabular}{lll}
\multicolumn{3}{c}{\bf LAD} \\
\hline
Parameter & Requirement & Goal \\
\hline
Energy range & 2--30~keV (nominal) & 1--40 keV (nominal) \\
             & 2--50~keV (expanded)& 1--60 keV (expanded) \\
\vspace{-0.2cm}\\
Eff. area & 12.0~m$^2$ (2--10 keV) & 15~m$^2$ (2--10 keV) \\
          & 1.3~m$^2$ (@30 keV) & 2.5~m$^2$ (@30 keV) \\
\vspace{-0.2cm}\\
$\Delta$E & $<$260 eV & $<$180 eV \\
(FWHM, @6 keV)  &  &  \\
\vspace{-0.2cm}\\
FoV (FWHM) & $<$60 arcmin & $<$30 arcmin \\
\vspace{-0.2cm}\\
Time res. & 10 $\mu$s & 5 $\mu$s \\
\vspace{-0.2cm}\\
Dead time & $<$0.5\% (@1 Crab) & $<$ 0.1\% (@1 Crab) \\
\vspace{-0.2cm}\\
Background flux & $<$10 mCrab & $<$ 5 mCrab \\
\vspace{-0.2cm}\\
Max. src flux (steady) & $>$0.3 Crab &  $>$10 Crab \\
\vspace{-0.2cm}\\
Max. src flux (peak) & $>$15 Crab &  $>$100 Crab \\
\hline
\vspace{0.2cm}\\
\multicolumn{3}{c}{\bf WFM} \\
\hline
Parameter & Requirement & Goal \\
\hline
Energy range & 2--50~keV  & 1--50 keV  \\
\vspace{-0.2cm}\\
$\Delta$E (FWHM)& $<$300 eV & $<$200 eV \\
\vspace{-0.2cm}\\
FoV (FWHM) & $>$3 steradian & $>$4 steradian \\
\vspace{-0.2cm}\\
Ang. res. & 5 arcmin & 3 arcmin \\
\vspace{-0.2cm}\\
PSLA & 1 arcmin & 0.5 arcmin \\
\vspace{-0.2cm}\\
Sens. (5$\sigma$, 50 ks) & 2 mCrab & 1 mCrab \\
\vspace{-0.2cm}\\
Sens. (5$\sigma$, 1 s) & 0.5 Crab & 0.2 Crab \\
\hline
\end{tabular}
\caption{Scientific requirements for the LOFT LAD and WFM instruments.}
\label{t:scireq}
\end{center}
\end{table}

The scientific requirements of the LOFT payload as derived by the
science objectives are summarized in Table~\ref{t:scireq}. The key
system requirements of the LOFT mission to meet such scientific
requirements are the accommodation of the structure hosting the
Large Area Detector (LAD), having a $\sim$20~m$^{2}$ surface area,
and the powering of such a large detector ensemble. A preliminary
evaluation by Thales Alenia Space Italy (TAS-I), on behalf of
INAF, has identified a mission concept combining such a large
structure with a standard service module. By means of deployment
mechanism technologies derived from SAR missions, like the ESA's
SMOS \cite{Plaza2003}, the satellite can be stowed within the
launcher fairing volume of a Vega, the small ESA launcher.
Considering a launch into a Low Equatorial Orbit (LEO,
$\sim$600~km) and the estimated mass of the flight segment, the
Vega launcher still provides a fairly large mass margin.

Although LOFT/LAD observations are largely source-dominated in
most cases, the science objectives related to the AGNs are very
sensitive to the issue of the background minimization and
stability, highly benefiting from the choice of a circular,
equatorial  ($<$5$^{\circ}$ inclination) LEO. The latter will be
also very effective in minimizing the long-term radiation effects
on the detectors.

LOFT has a higher data throughput compared to previous X-ray
missions. We envisage a telemetry downlink to the Kourou (ESA) and
also to the Malindi (ASI) equatorial ground stations. In this
scenario each ground station will have a $\sim$10-min contact
every $\sim$100-min orbit. Kourou will be used for both
telecommand uplink and telemetry downlink. Malindi for telemetry
downlink only. The maximum rate available to the scientific
telemetry (orbit average) in this configuration is $\sim$700 kbps.

The LOFT mission is expected to operate in three-axes stabilized
pointing mode. The slew rate capability is estimated in about 4$^{\circ}$/minute. 
The solar panel array is fixed, and the Sun aspect
angle constraint is set at $\pm$20$^{\circ}$. An orbital
simulation has been performed considering a launch in 2022 (solar
minimum period) and ballistic coefficient based on the nominal
satellite attitude. The simulation shows an orbital decay of
$\sim$20~km in a period of 5 years. Based on these preliminary
evaluations it is possible to assert the feasibility of LOFT
mission as a small-medium class mission in LEO equatorial orbit,
with a Vega launch vehicle.

\section{Model Payload}\label{s:model_payload}
\begin{figure}[t]
\centering
    \includegraphics[width=0.6\textwidth]{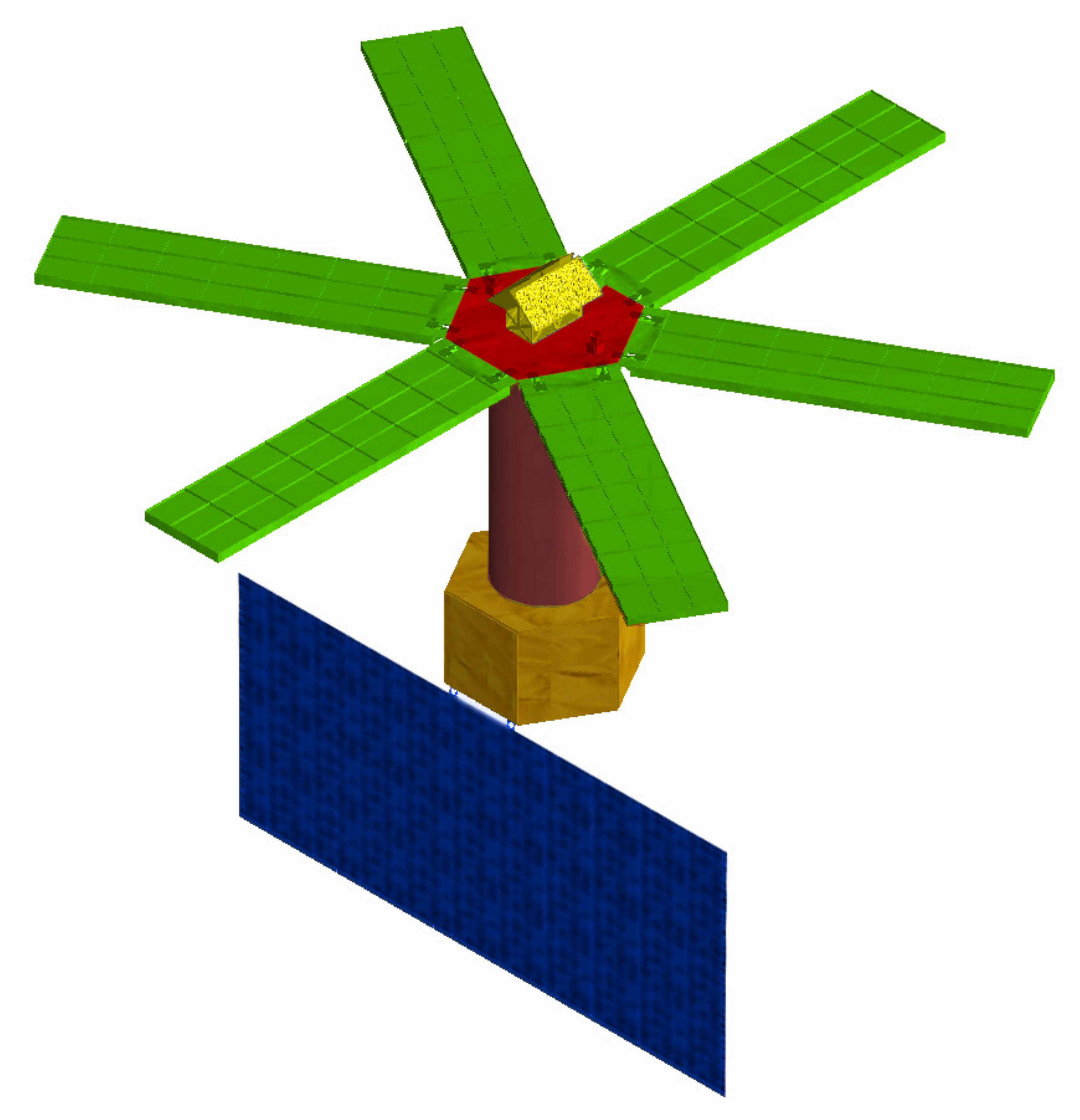}
    \caption{Conceptual scheme of the LOFT satellite. (green) Large
      Area Detector, (yellow) Wide Field Monitor, (red) Optical Bench,
    (purple) Structural Tower, (gold) Bus and (blue) Solar Array.}
    \label{f:LOFT_conceptual}
\end{figure}

To meet the scientific requirements summarized in
Table~\ref{t:scireq} and to achieve the science objectives
summarized in Sect.~\ref{s:scientific_objectives}, the LOFT
scientific payload is composed of two instruments: the Large Area
Detector (LAD) and the Wide Field Monitor (WFM). Both experiments
rely on the technology of the large-area Silicon Drift Detectors
(SDDs), although with application-specific detector design
details. In Fig.~\ref{f:LOFT_conceptual} the different components
of the LOFT satellite are identified: the LAD is composed of 6
Detector Panels (DPs) deploying from an optical bench, hosting the
WFM at its center, observing a sky region including the field of
view of the LAD.

\subsection{The Large Area Detector (LAD)}
The LAD is a collimated experiment, conceptually similar to its
predecessors (e.g. RXTE/PCA, \cite{Jahoda2006}). The development
of a 10~m$^2$-class experiment is now made possible by the recent
advancements in the field of large-area silicon detectors - able
to time tag a X-ray photon with an accuracy $<$10~$\mu$s and an
energy resolution of $\sim$250~eV FWHM - and capillary-plate X-ray
collimators. The relevant feature of the LOFT design is the low
mass and power per unit area enabled by these two elements. In
this respect, the key properties of the Si drift detectors (see
e.g. \cite{Gatti1984}) are their capability to read-out a large
photon collecting area with a small set of low-capacitance (thus
low-noise) anodes and their very small weight
($\sim$1~kg~m$^{-2}$).

\subsubsection{The detector}\label{s:detector}
The primary enabling technology for the LAD is the large-area
Silicon Drift Detectors (SDDs) based on the heritage of the
detectors developed for the Inner Tracking System (ITS) in the
ALICE experiment of the Large Hadron Collider (LHC) at CERN, by
INFN Trieste, Italy - in co-operation with Canberra Inc.
\cite{Vacchi1991, Rashevsky2002}.

The LAD detector design is indeed an optimization of the ALICE
detector: 6-inch, 450~$\mu$m thick wafers will be used to produce
76~cm$^2$ monolithic SDDs (108.52~mm$\times$70.00~mm active area).
The anode pitch is increased to 854~$\mu$m (corresponding to an
elemental area of 0.854~mm$\times$35~mm = 0.299~cm$^2$) to reduce
the power consumption and improve the low-energy response. The Si
tile is electrically divided in two halves, with 2 series of 128
read-out anodes at two edges and the highest voltage along its
symmetry axis. The drift length is 35~mm. A drift field of
370~V/cm (1300~V maximum voltage), gives a maximum drift time of
$\sim$5~$\mu$s \cite{Kushpil2006} at room temperature, constituting the detector
contribution to the uncertainty in the determination of the time
of arrival of the photon. Indeed, with the above electric field,
the charge will typically distribute over 1 (40\% of events) or 2
(59\% of events) anodes, rarely on 3 (1\% of counts). The
segmentation of the detector makes the dead time and pile-up risks
negligible (see below), a crucial property for a timing
experiment.

\begin{figure}[t]
\centering
    \includegraphics[width=0.5\textwidth]{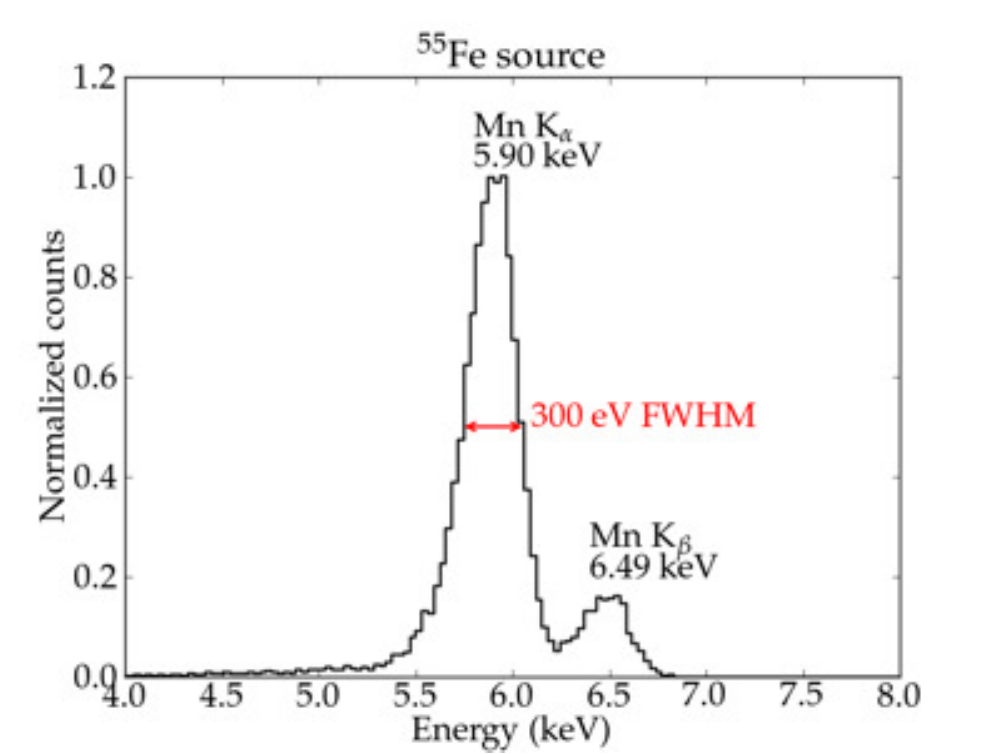}
    \caption{Energy spectra measured using a spare ALICE detector equipped with discrete read-out electronics, at room temperature. The FWHM energy resolution was measured as $\sim$300~eV at 5.9~keV. }
    \label{f:fe55_spectrum}
\end{figure}

In Fig.~\ref{f:fe55_spectrum} we show examples of energy spectra of a $^{55}$Fe source (with
lines at 5.9 and 6.5~keV) obtained at 20$^{\circ}$C, by instrumenting a
spare ALICE SDD with discrete electronics. The energy resolution is
$\sim$300~eV FWHM on single-anode events and the low energy
discrimination threshold is below 0.6~keV (see \cite{Zampa2011} for details).

The detectors as optimized for the LOFT application will achieve
even better performance. In fact, the current performance of the
prototype breadboard at room temperature (see \cite{Zampa2011})
will be improved by the use of an integrated read-out electronics
that will largely reduce the parasitic capacitance, and by the
instrument operation at low temperature, decreasing the leakage
current by more than an order of magnitude, even in the larger
thickness and pitch configuration of LOFT, as well as with the
expected level of radiation damage in orbit.

\subsubsection{The collimator}\label{s:collimator}
The other key element of the LOFT payload innovative design is the
capillary-plate X-ray collimator (although collimators based on
the same technology were already used in the Medium Energy
experiment onboard EXOSAT, \cite{Taylor1981}). This is based on
the technology of micro-channel plates. A multi-pore, $\sim$mm
thin sheet of lead-glass is able to absorb soft X-rays coming from
outside its aperture holes. Standard collimators can be procured
off-the-shelf with a 50\% lead content in the glass, offering a
significant stopping power to soft X-rays. We studied the angular
response to different photon energies with GEANT Monte Carlo
simulations and the results demonstrate an effective shielding up
to $\sim$40~keV, above the energy range of LOFT. Photons with
higher energy passing through will be detected with low efficiency
by the thin Si detector and will be anyway discriminated by their
energy deposition, except for the Compton interactions that will
contribute to the residual instrumental background. A noticeable
property of these devices is their very low weight. The LOFT
baseline is a 2~mm thick plate, with inner hole diameter of
25~$\mu$m and pitch of 28~$\mu$m (field of view 43~arcmin FWHM,
OAR=80\%): the net weight is about 3~kg~m$^{-2}$ enabling large
areas at reasonable mass and costs.

\subsubsection{The Front-End and Back-End Electronics}\label{s:fee}
The requirements of fine pitch, small parasitic capacitance,
low-power consumption and the need for tiling imply a high-density
read-out system, based on ASICs (Application Specific Integrated
Circuits). The low capacitance of each anode allows us to reach
ultra low noise levels  and short optimum signal processing
time ($\simeq$1~$\mu$s). The specifications for the baseline LOFT
ASIC, developed by Polytechnic of Milan and University of Pavia
 include: an input range of 0.5--100~keV
(135--27000~e$^{-}$); a noise $\leq$ 20~e$^{-}_\mathrm{rms}$;
12~bit ADC resolution; a time tagging accuracy of an event to
better than 1~$\mu$s and a power consumption $<$540~$\mu$W/channel
(all inclusive).

The event processing time in the ASIC drives the dead time of the
system. Even in the worst case of an event involving 3 anodes
(less than 1\% of the cases), the corresponding detector area
being 0.9~cm$^2$, the SDD read-out time $<$10~$\mu$s makes the
pile-up probability negligible ($<$10$^{-4}$ for a source with a flux lower than 5 Crab). 
Based on the ASIC architecture, nearly simultaneous triggers on different channels
will be processed independently. The source count rate during the
Crab observation will then cause a dead-time $<$0.03\%.

A back-end electronics (BEE) is in charge of interfacing the
front-end electronics (FEE) with the Payload Data Handling Unit
(PDHU). It is organized hierarchically, with one Module Back-End
Electronics (MBEE) for each of the 21 Modules in a Detector Panel
(DP), connected to one Panel Back-End Electronics (PBEE) for each
of the 6 DPs. The full LAD will then have 126 MBEEs and 6 PBEEs.
The tasks of the BEE will include: power supply, post-regulation
and filtering; interfacing the digital output of the 16 FEEs (each
one including 8 ASICs) of each Module, as well as the PDHU; event
trigger filtering and time tagging; pedestal subtraction; cluster
identification and validation.

\subsubsection{The LAD Detector Panel organization}
\begin{figure}[!t]
 \centering
  \subfigure{ \includegraphics[height=0.31\textwidth, angle=90]{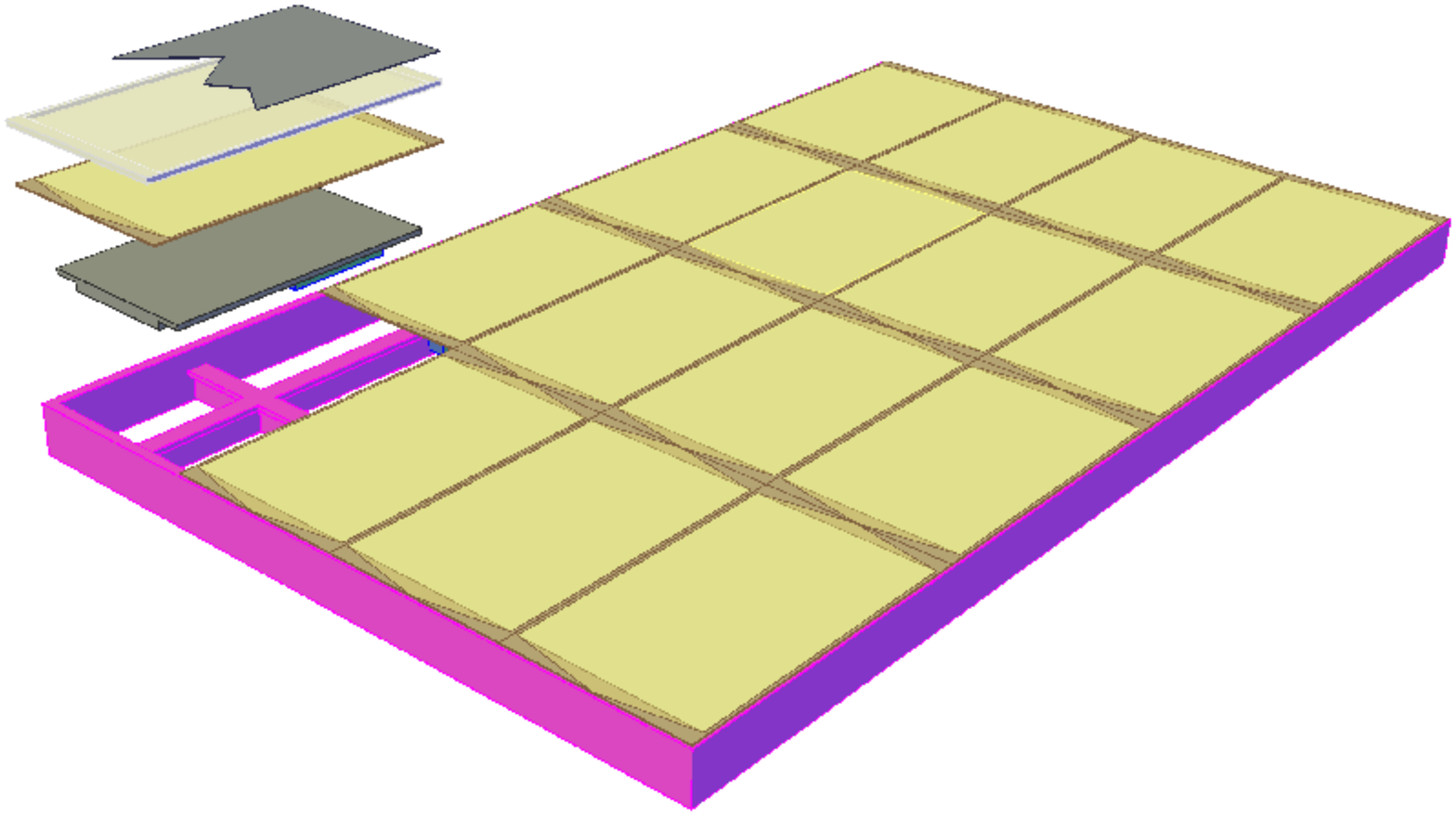} }
  \subfigure{ \includegraphics[height=0.31\textwidth, angle=90]{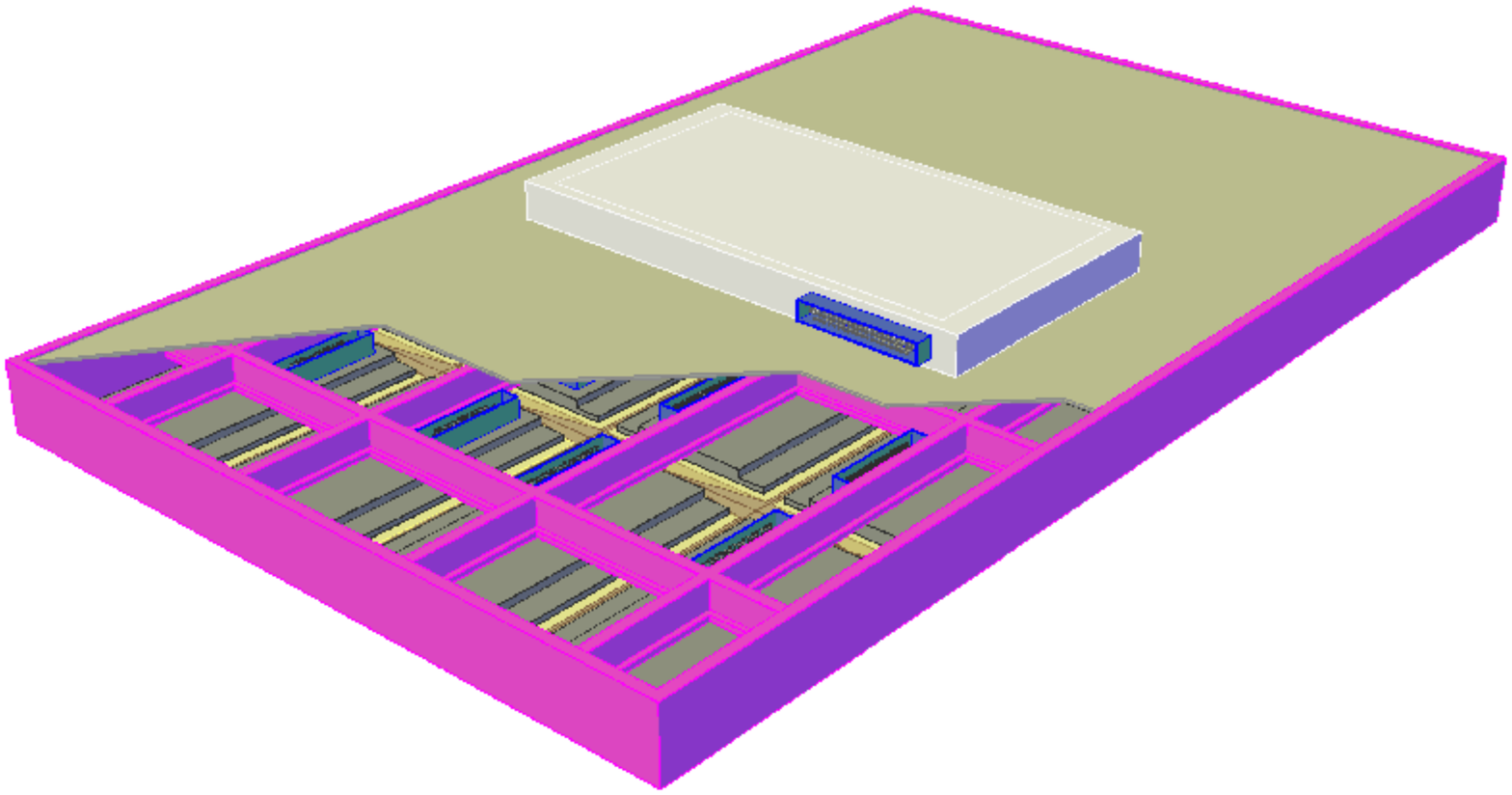} }
  \subfigure{ \includegraphics[width=0.31\textwidth, angle=5]{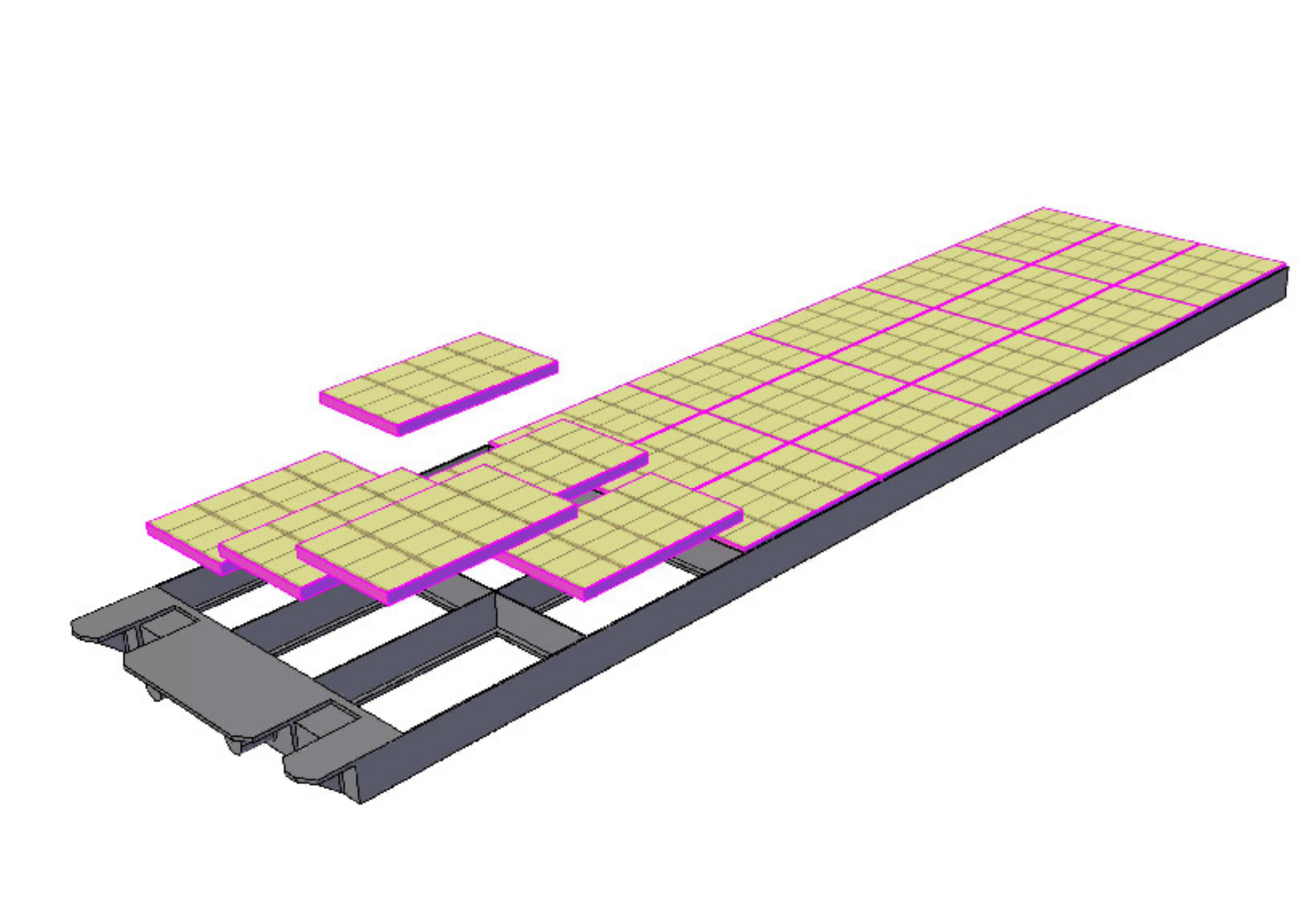}}
   \caption{\emph{Left:} One LOFT Module (top view). \emph{Right:} One LOFT Module (bottom view). \emph{Below:} A LOFT Detector Panel. }
\label{f:assembly}
\end{figure}

The basic LAD detection element is composed of SDD, FEE and
collimator. The assembly philosophy envisages a hierarchical
approach: Detector, Module, Panel, LAD Assembly. To provide an
efficient tiling, with the required mechanical stiffness and
alignment accuracy, the procedure for the Module integration
includes a reference frame where the 16 collimators in a Module
can be accurately integrated with the detector and the FEE
(Fig.~\ref{f:assembly}). The latter is composed of a thin printed
circuit board, designed to host the SDDs on one side and the ASIC
and electronics component on the opposite side, with pass-throughs
for electrical connections. The power lines to the voltage divider
(HV and MV) are brought from the back (p-side) to the front
(n-side) pads of the Si tile by a wrap-around flex cable, similar
to that designed for the ALICE experiment \cite{Nouais2003}.

The SDDs are sensitive to visible/UV light, which causes an
increase in the leakage current. To prevent this, each detector
will be equipped with a multilayer blanket, operating both as a
light and thermal shield. We selected 1~$\mu$m thick Polymide film
coated with 0.08~$\mu$m Aluminum. This choice will ensure an
optical transmission below 10$^{-6}$, and a soft X-ray
transmission greater than 90\% at 2~keV.

The back side of the Module (Fig.~\ref{f:assembly}, center) will
also host the mechanical interface to the Module Back-End
Electronics (MBEE), as well as a shield to stop photons from the
cosmic X-ray background. The surface of the MBEE will also be
designed as a radiator. The power generation in a Module is evenly
shared between MBEE and detectors: a larger detector radiator will
favor a lower temperature on the SDDs and ASICs, improving the
performance, while keeping the rest of the electronics at moderate
cooling.
A set of 21 Modules will be integrated into a Detector Panel (DP,
Fig.~\ref{f:assembly}). This is a mechanical frame providing
mechanical support and alignment interfaces for the individual
Modules, for the Panel Back-End Electronics (PBEE) box and
enabling the electrical connection and routing between the 21
MBEEs and the single PBEE in each DP. The DP will have mechanical
and alignment interfaces with the LAD deployment mechanism.

The LAD deployment mechanism will allow the LAD to be launched in
a stowed position that fits within the Vega fairing. The basic
mechanism draws on the heritage of Synthetic Aperture Radar (SAR)
panel deployment systems. This type of mechanisms guarantees
accuracy and repeatability in the panel deployment in the
sub-arcmin range (e.g., \cite{Bueno2005}). The preliminary LOFT
satellite configuration consists of 6 DPs, connected to the
satellite optical bench by hinges. The mechanisms will provide the
deployment with a pure mechanical control and motorization.
\begin{figure}[!t]
 \centering
  \subfigure{ \includegraphics[width=0.30\textwidth]{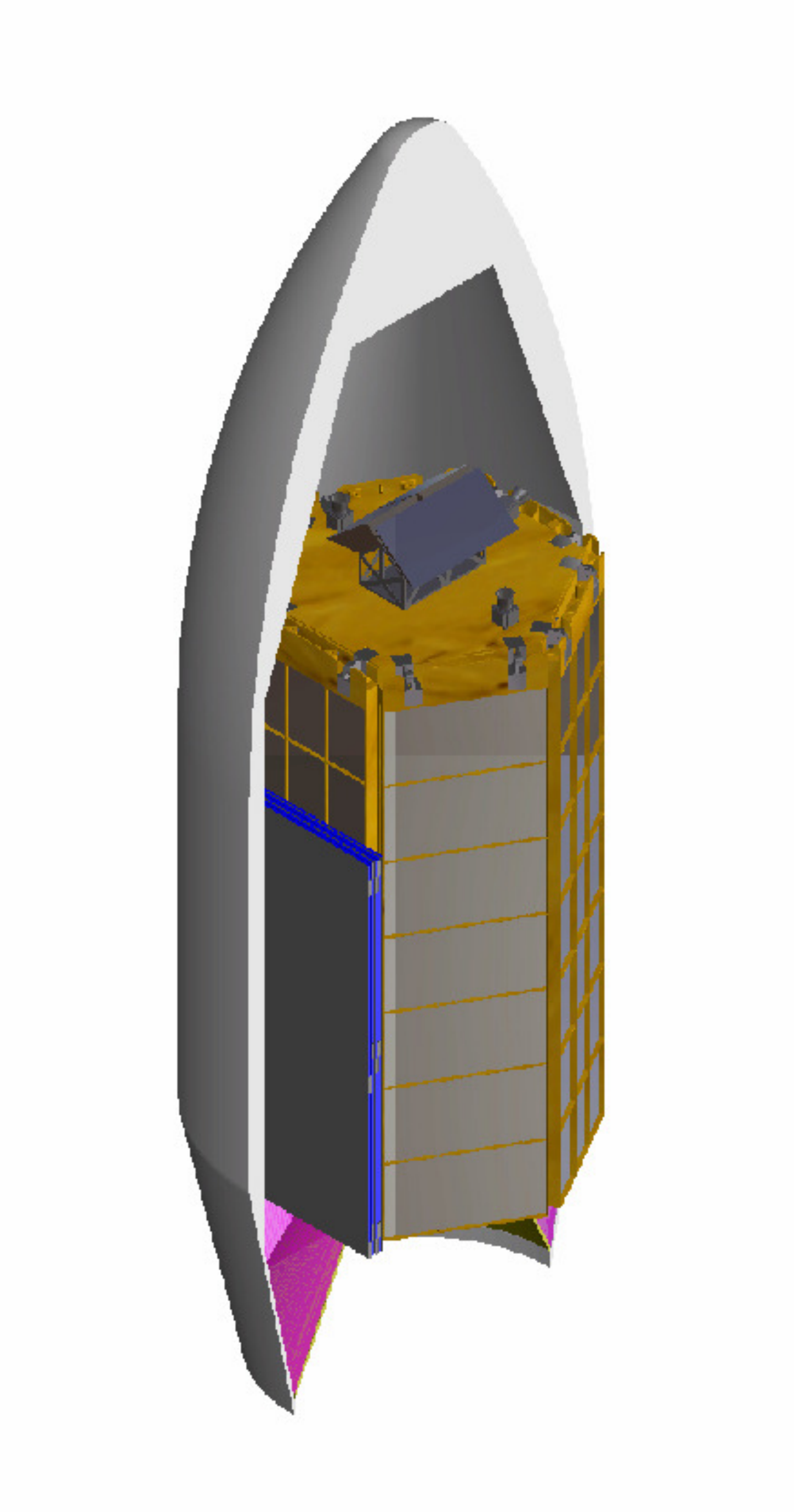} }
  \subfigure{ \includegraphics[width=0.60\textwidth]{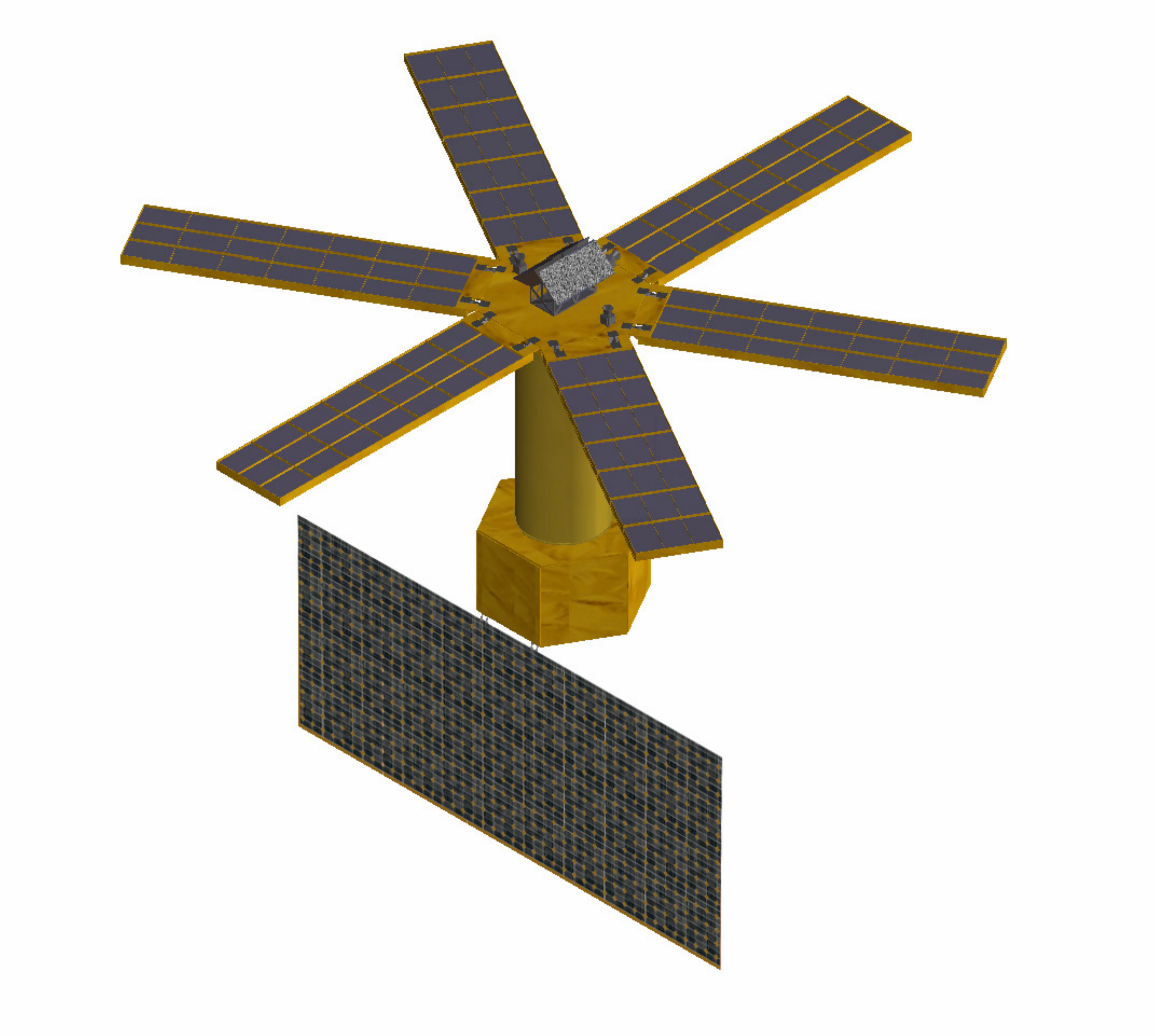} }
   \caption{The LOFT satellite inside the Vega launcher (left) and
     after the deployment of the LAD and of the solar array (right).}
\label{f:stowed_and_deployed}
\end{figure}

The LAD experiment has a favorable structure for temperature
control, due to the uniform power generation over the DPs, that
can be used in the thermal balance with irradiation for a passive
thermal control. Indeed, a preliminary thermal analysis shows that
the LOFT detector panels are expected to operate at temperatures
between -30$^{\circ}$C and 0$^{\circ}$C, depending on the Sun
aspect angle and the thermal design, with a short-term (e.g.,
orbital) variation of less than $\sim$3$^{\circ}$C. A design
optimization will be performed to achieve a low and stable
operating temperature, needed for an optimal detector and FEE
performance.

\subsubsection{Mass, Power, and Telemetry budgets}\label{s:budgets}
Based on the preliminary design described in the previous
sections, we evaluated detailed mass and power budgets. Our
evaluation shows an overall satellite mass of $\sim$1800~kg,
including both sub-system and system level margins, with
approximately 60\% allocated to the payload. The power budget was
also computed on the basis of detailed subsystem estimation. The
total power required by the satellite, including all the subsystem
and system margins, is $\sim$1800~W, compliant with the available
power generated by a solar panel array that can fit the satellite
and launch constraints.

The LAD telemetry budget is estimated under the following
assumptions: $i.$ default event-by-event data transmission, $ii.$
individual event info: $\sim$60~eV energy bin below 30~keV,
$\sim$2~keV in 30--50~keV (expanded range), 2~$\mu$s time
resolution and $iii.$ absolute time event every 100~ms, and
differential event time tagging. In estimating the telemetry
budget, we assume a source with intensity 500 mCrab permanently in
the field of view (this flux threshold includes $>$95\% of the
known X-ray sources with flux above 1~mCrab, \cite{Ebisawa2003}).
The overall LAD telemetry budget is $\sim$600~kbps, through a
lossless compression algorithm (factor of $\sim$3-4).
 A mass memory on the PDHU will allow the temporary storage
of excess telemetry. Some of the key science targets ($\sim$10
persistent sources and some bright X-ray transients) will exceed
the above flux threshold assumption. In these cases we will employ
a flexible set of data modes, as was done with the Event Data
System (EDS) on RXTE. These modes allow the time and energy
binning to be optimized for the science goals within the available
telemetry budget.

\subsubsection{The LAD effective area}\label{s:lad_performance}
\begin{figure}[h]
\centering \subfigure{
\includegraphics[width=0.7\textwidth]{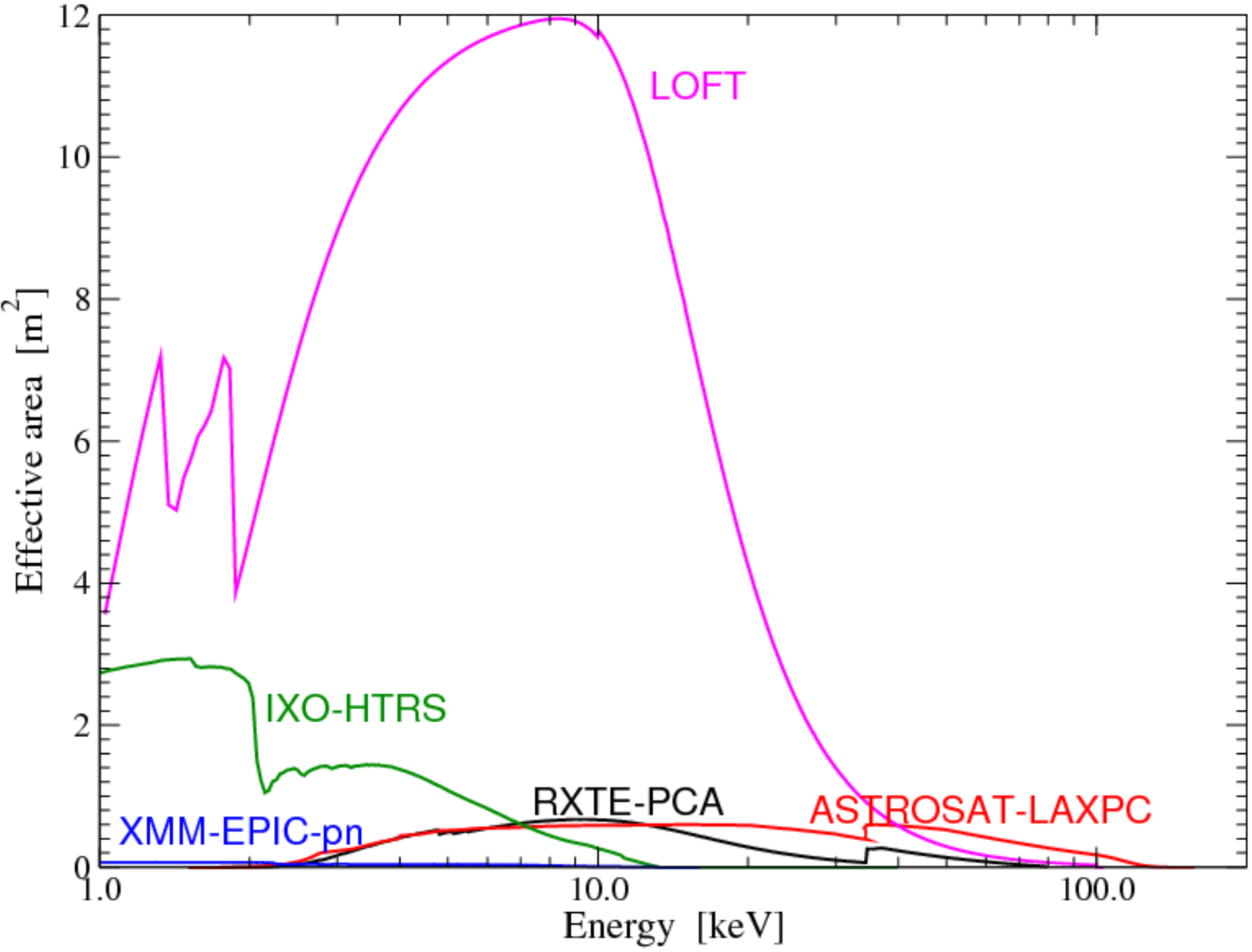}}
\subfigure{
\includegraphics[width=0.7\textwidth]{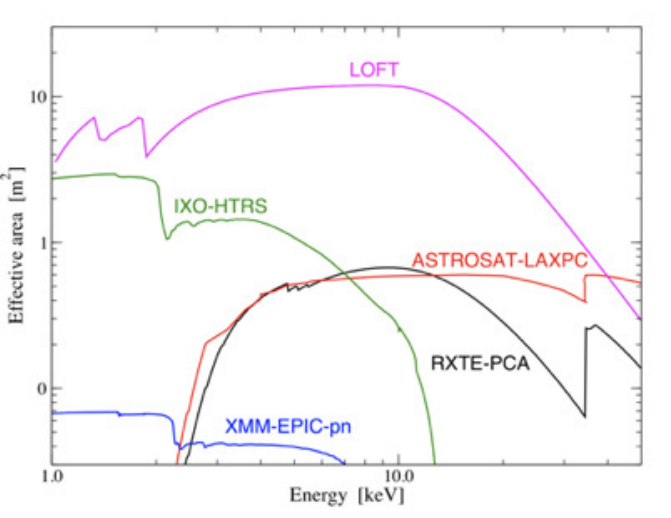}}
\caption{LAD effective area (vs. energy) plotted in both linear
and logarithmic scale, as compared to that of other satellites for
X-ray astronomy.} \label{f:LAD_eff_area}
\end{figure}

The preliminary design described in the previous sections allows
to obtain an effective area for the LAD as shown in
Fig.~\ref{f:LAD_eff_area}, as a function of energy. In the same
plot the effective area of timing experiments onboard previous and
planned missions is shown for comparison. The plot is shown in
both log-log and log-lin scale, as the sensitivity for the QPOs
scales linearly with the count rate, that is with the effective
area: the linear vertical scale allows to appreciate the
breakthrough offered by LOFT in this field.

The energy dependence is driven by the detector properties. The
effective area peaks at a value of $\sim$12~m$^{2}$ in
$\sim$5-10~keV. The factor primarily affecting the resulting peak
effective area with respect to the $\sim$15~m$^{2}$ geometric area
is the open area ratio of the collimator.

The low energy decrease is due to the passive materials
intervening in the detector field of view (including both the
detector passive layers and electrodes, as well as the
optical-thermal blanket). In the energy range between 1 and 2~keV
an effective area as large as $\sim$5 m$^{2}$ is still available
in principle. However, the current baseline for the low energy
threshold is set at 2~keV: a lower energy threshold is a goal and
it is currently being investigated. At high energies the decrease
of the effective area is due to the decreasing quantum efficiency
of the 450~$\mu$m thick silicon. However, a noticeable value of
$\sim$1.3~m$^{2}$ for the effective area is still available at 30
keV, significantly larger than any other timing experiment
(Fig.~\ref{f:LAD_eff_area}). Improvements at higher energies
require thicker detectors, that are currently under study.

\subsection{The Wide Field Monitor (WFM)}
\begin{figure}[t]
 \centering
 \subfigure{ \includegraphics[width=0.45\textwidth]{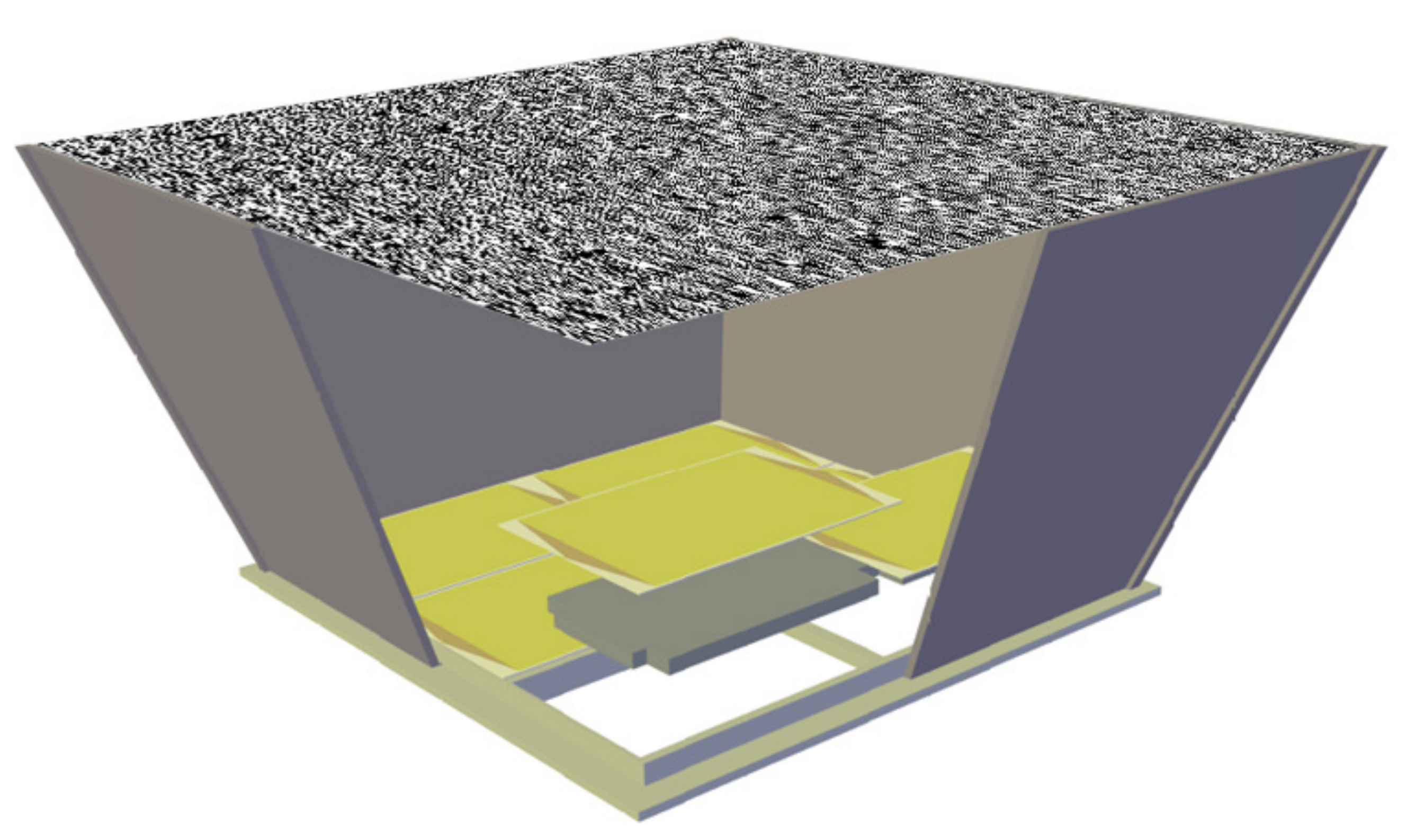} }
  \subfigure{ \includegraphics[width=0.45\textwidth]{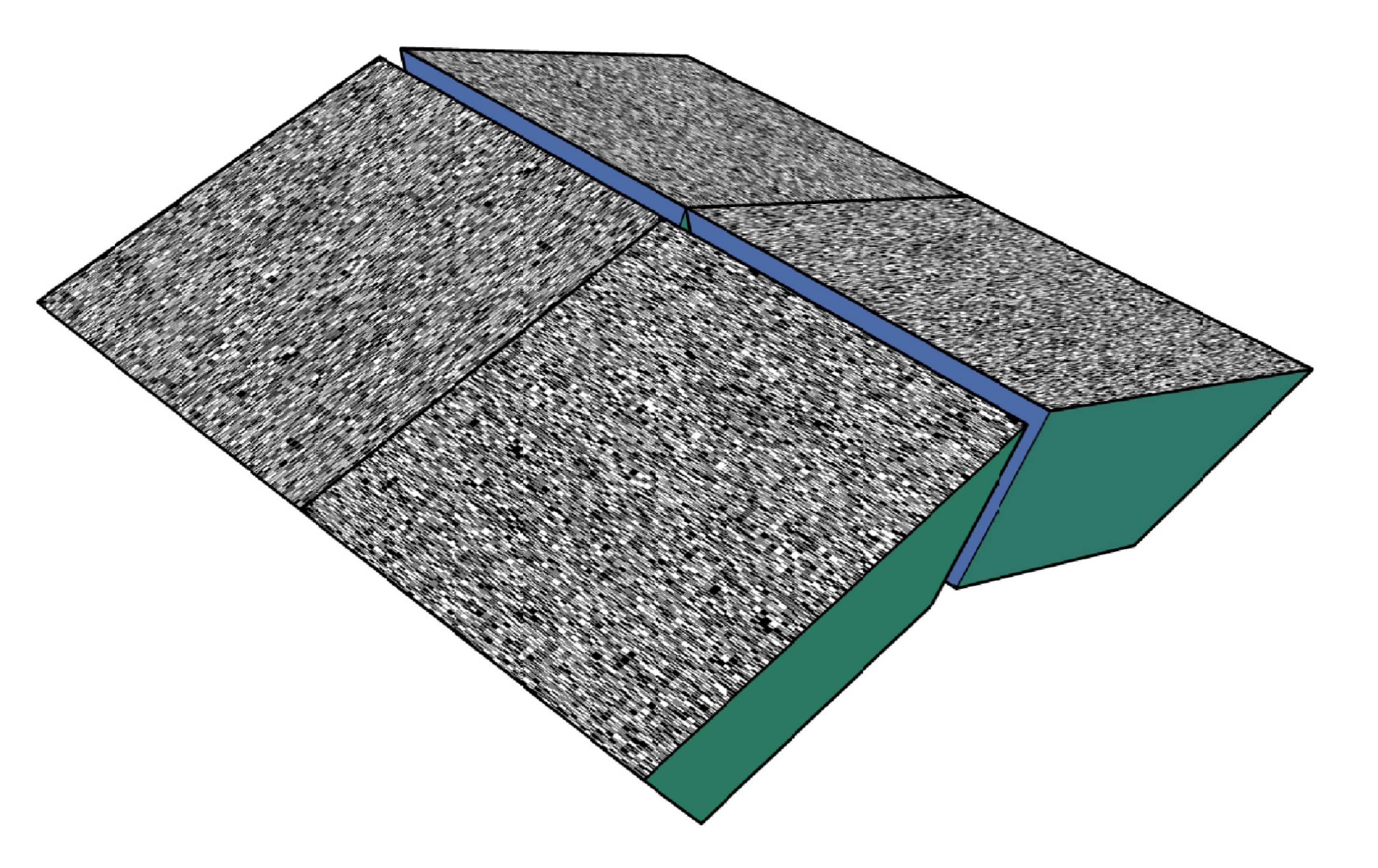} }
   \caption{The Wide Field Monitor.}
\label{f:WFM}
\end{figure}
The scientific requirements of the WFM are reported in
Tab.~\ref{t:scireq}. The WFM is composed of 4 coded mask cameras
for a total geometric area of 1600 cm$^{2}$, hosted on the top of
the LAD tower. It is designed on the heritage of the SuperAGILE
experiment \cite{Feroci2007}, successfully operating in orbit
since 2007 (e.g., \cite{Feroci2010}). The latter demonstrated the
feasibility of a compact, large-area, light and low-power, arc
minute resolution X-ray imager, with steradian-wide field of view.
The LOFT WFM applies the same concept, with improvements provided
by the superior performance of the large area SDDs (similar to
those in the LAD, but with a finer anode pitch) in place of the
silicon microstrips. The different detectors and their
configuration allow the LOFT/WFM to operate down to much lower
energies (2 keV).

By using the large area SDDs, with a position resolution
$<$100~$\mu$m, a coded mask at $\sim$150~mm provides an angular
resolution $<$5~arc~min. The coded mask imaging is the most
effective technique to observe simultaneously steradian-wide sky
regions with arc min angular resolution. In contrast to their use
in the LAD, in the WFM the position resolution of the SDD is a key
issue. For this reason the baseline WFM SDDs have a 
$\leq$300~$\mu$m anode pitch. Using the lab tests
of the ALICE spare detector (see Sect.~\ref{s:detector}) and an accurate
Monte Carlo simulation of the detector physics, we estimated a
position resolution better than 60~$\mu$m FWHM along the anodes and
$\sim$8~mm FWHM along the drift channel \cite{Campana2011}. The latter
exploits the electron cloud diffusion during the drift that makes
the charge distribution over contiguous anodes dependent on the
drift distance.

\begin{figure}[ht]
 \centering
  \includegraphics[height=0.4\textwidth]{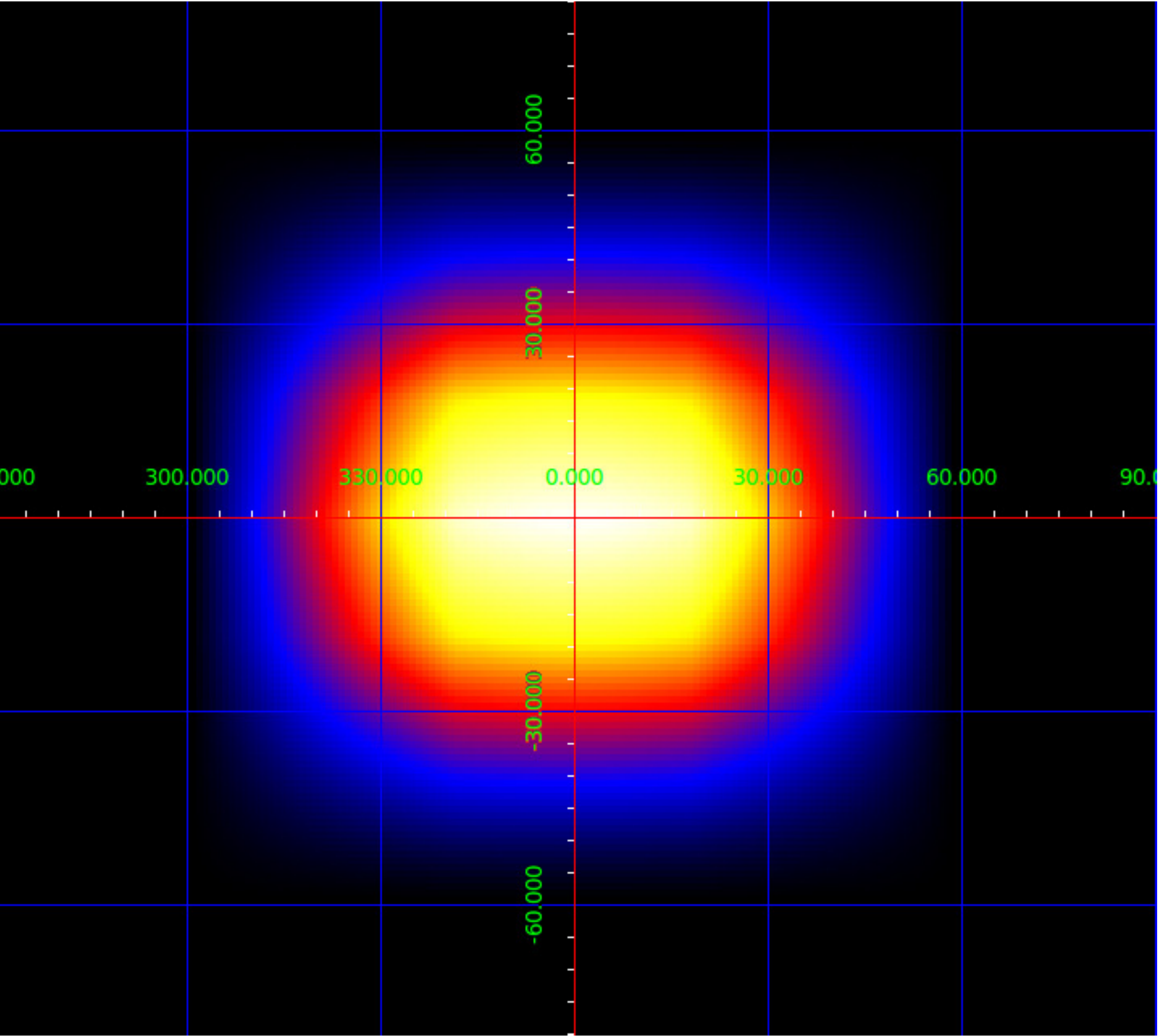}
  \includegraphics[height=0.4\textwidth]{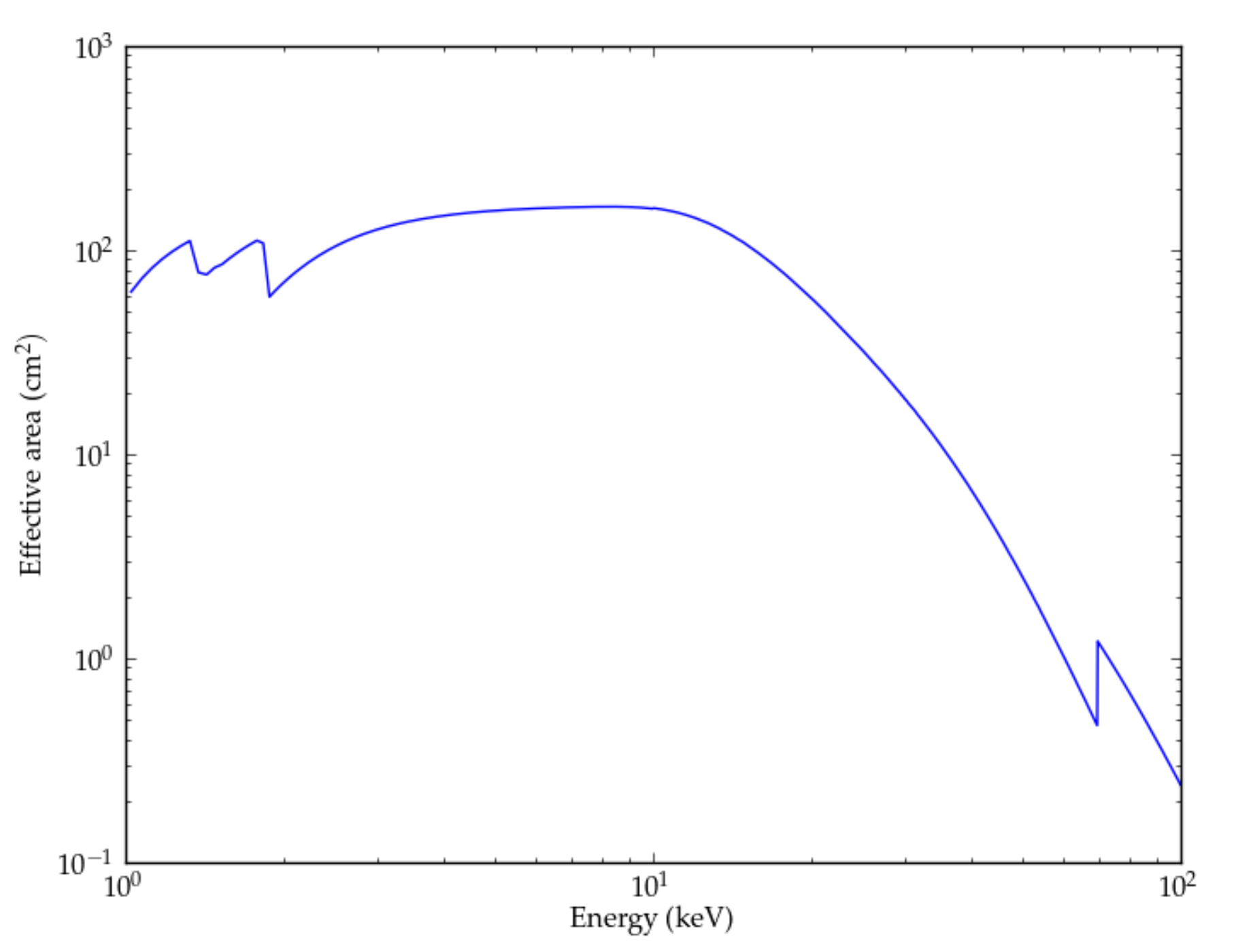}
   \caption{Field of view of one of the four WFM units, and effective area as a function of energy.} \label{f:WFM_fov}
\end{figure}

Due to the coarse detector position resolution along the drift
direction, we conservatively consider the WFM camera as 1-D and we
require 2 orthogonal cameras to image the same sky region at any
time for fine source positioning in 2-D. This is the same
technique used in SuperAGILE, RossiXTE/ASM or HETE/WFM. However,
in the LOFT/WFM the coarse position resolution in the drift
direction in each camera reduces
 considerably the source confusion in crowded fields (e.g., Galactic
Center). The design of a single WFM camera (Fig.~\ref{f:WFM})
consists of a tiled 20~cm$\times$20~cm detector plane (6
large-area 450~$\mu$m thick SDDs), an asymmetric 2-D coded mask
and a collimator. The SDD read-out employs the same ASICs as the
LAD, with a proper design customization to the finer pitch. A
light collimator (e.g., 1~mm thick carbon fibre, covered with a
150~$\mu$m tungsten sheet) effectively shields the X-ray photons
from the diffuse X-ray background (the main background source, as
charged particles are efficiently rejected by the amplitude
discrimination) up to energies above 60~keV. The 150~$\mu$m thick
tungsten mask (Fig.~\ref{f:WFM}) has a pitch of $\sim$250~$\mu$m
and $\sim$16~mm in the two directions, respectively. The mask size
is 30~cm$\times$30~cm, in order to achieve a ``flat response''
(i.e., equally sensitive, see Fig.~\ref{f:WFM_fov}) over the
central part of the field of view.

As a baseline, due to telemetry limitations, the WFM will not
operate in photon-by-photon mode at all times. Images in different
energy ranges every 300 s, 0.1 keV resolution energy spectra every
30 s and light curves (bin size 16 ms) will be integrated and
pre-processed by the BEEs and PDHU onboard (similar to what
routinely done by Swift/BAT). A triggering algorithm will scan
time series and images to detect fast transients, e. g., due to
X-ray bursts or GRBs. Upon trigger, $\sim$300 s of photon-by-photon 
data will be downloaded to the telemetry. This mode of
operation will provide the full achievement of the primary science
objective of the WFM: source monitoring and detection of
interesting intensity and spectral states. Should a larger
telemetry rate become available, photon data will be transmitted
to the ground, enhancing the spectral and timing capabilities of
this instrument.

The data processing requires an FPGA-based Back-End Electronics
(BEE, one individual BEE for an orthogonal pair of units, as a
redundancy) for the power DC/DC conversion and filtering, event
cluster analysis, pedestal calculation and subtraction, time
tagging, housekeeping, etc. The WFM total power consumption is
estimated, including margin, as 12~W.

The telemetry packets of the WFM contain images, light-curves and
energy spectra, accumulated on-board. The PDHU will reconstruct
the event position and energy over the detector using the
individual anode amplitude information, through proper algorithms
and look-up tables. Upon trigger, the WFM data are sent to ground
in photon-by-photon mode in dedicated telemetry packets. Total
telemetry rate is estimated as 91~kbps (orbit average, before
compression), including an average trigger rate of 1/day.

The overall WFM design includes 4 identical camera units assembled
in a configuration where an orthogonal set of 2 cameras monitors
the same sky region. The two sets are oriented at 37$^{\circ}$ to
each other. The anticipated characteristics and performance of
this WFM design and configuration are given in
Table~\ref{t:WMF_performance}.

The WFM total volume is 4$\times$(30$\times$30$\times$15~cm$^3$),
while the total mass, including 20\% margin, is estimated as
37.0~kg.
\begin{table}
\begin{center}
\begin{tabular}{lcc}
\multicolumn{3}{c}{\bf }\\
\vspace{-0.1cm}\\
\hline
Parameter & Single Unit & Overall WFM \\
\hline
Energy range & 2--50~keV (nominal) & 2--50 keV (nominal) \\
\vspace{-0.2cm}\\
Geometric Area & 400 cm$^2$ & 1600~cm$^2$ \\
\vspace{-0.2cm}\\
Energy Resolution FWHM & $<$350~eV &  $<$350~eV  \\
\vspace{-0.2cm}\\
FoV fully coded & 0.40~sr & 0.80~sr \\
FoV partially coded & 2.90~sr & 3.95~sr \\
Zero Response & 118$^{\circ}$ & 154$^{\circ}$ \\
\vspace{-0.2cm}\\
Angular Resolution FWHM & $6'\times7^{\circ}$ & $6'\times6'$  \\
\vspace{-0.2cm}\\
Source Location Accuracy (10$\sigma$) & $<$ $1' \times 40'$ & $<$ $1' \times 1'$ \\
\vspace{-0.2cm}\\
On-axis sensitivity at 5$\sigma$ in 1~s & 640~mCrab & 450~mCrab \\
\vspace{-0.2cm}\\
On-axis sensitivity at 5$\sigma$ in 50~ks & 2.9~mCrab & 2.0~mCrab \\
\vspace{-0.2cm}\\
\hline
\end{tabular}
\caption{Performance of a single unit of the WFM and of the
overall WFM}
\label{t:WMF_performance}
\end{center}
\end{table}

\section{LOFT Observing strategy and data flow}
The LOFT mission will be operated by ESA as an observatory open to
the general scientific community. The MOC and SOC are foreseen to
be run by the ESA centers, while the science data center will be
proposed at the ISDC (Geneve). The latter will perform prompt and
automated data screening in order to guarantee a proper prompt
reaction to interesting transient events and source states.

The typical observing slots will last from a few ks to a few days,
depending on the target and strategy. Given the characteristics of
the LOFT scientific investigations, target of opportunity
observations will be carried out, as triggered by the LOFT/WFM or
by external triggers, in order to observe the variable sources in
their most interesting and extreme physical states.

\section{Summary and Conclusions}
The LOFT mission proposal was selected by ESA as one of the four
candidate M3 missions for the Cosmic Vision programme. ESA and the
LOFT consortium will carry out an assessment phase of the mission
over $\sim$18 months, concluding by the end of 2012. Further
down-selections of the mission candidates are currently expected
to bring only one of the four M3 candidates to a launch in the
time-frame 2020-2022.

LOFT was primarily designed to address the fundamental questions
raised by the Cosmic Vision under the "matter under extreme
conditions" theme, through the study of the spectral and fast flux
variability of compact X-ray sources. The high throughput required
by this type of investigation is achieved by an unprecedented
large collecting area: 15 m$^{2}$ geometric area, leading to
12~m$^{2}$ effective area peak at 6-10 keV (see
Fig.~\ref{f:LAD_eff_area}), 20 times larger than any predecessors.
The enabling technology is provided by the large area Silicon
Drift Detectors and the capillary plate X-ray collimators,
offering a large collecting and collimated area with limited
resources in terms of mass, volume, power and costs. This type of
detectors combine large area with very good spectral performance,
enabling spectral studies with a resolution of $\sim$250~eV (FWHM).
The segmentation of the Si drift detectors is a straightforward
solution to the issues of dead time and pile-up, among the main
problems in timing experiments.

Time variability from sub-ms to days will be explored by the LOFT
LAD and WFM with statistical accuracy never achieved to date. For
example, for the first time quasi-periodic oscillations in neutron
star or black-hole systems will be studied in the time domain (as
opposite to the frequency domain), within their coherence
timescale. The large effective area and the spectral capabilities
of LOFT will make optimal use of the diagnostic potential of the
fast variability of the X-ray emission to probe the behaviour of
matter in the strong gravity and ultradense environments, such as
the innermost stable orbits around black holes or the interior of
neutron stars, opening access to the understanding of the physical
laws at work in such extreme conditions.

\bibliographystyle{spmpsci}      
\bibliography{References}   
\end{document}